\documentclass[12pt,a4paper]{article}
\pdfoutput=1
\usepackage{jheppub}
\usepackage{mathtools,extarrows}

\usepackage{epstopdf}
\usepackage{graphicx}
\usepackage{ulem}
\usepackage{epsfig}
\usepackage{dcolumn}  
\usepackage{bm}    
\usepackage{amssymb} 
\usepackage{amsmath,bm}
\usepackage{amsfonts}    
\usepackage{slashed}  
\usepackage{youngtab}
\usepackage[mathscr]{euscript}
\usepackage{verbatim}
\usepackage[english]{babel}
\usepackage{color}
\usepackage{tikz}
\usepackage{braket}
\usepackage{subfigure}
\usepackage[colorinlistoftodos]{todonotes}
\usepackage{float}

\newdimen\tableauside\tableauside=1.0ex
\newdimen\tableaurule\tableaurule=0.4pt
\newdimen\tableaustep
\def\phantomhrule#1{\hbox{\vbox to0pt{\hrule height\tableaurule width#1\vss}}}
\def\phantomvrule#1{\vbox{\hbox to0pt{\vrule width\tableaurule height#1\hss}}}
\def\sqr{\vbox{%
                \phantomhrule\tableaustep
                \hbox{\phantomvrule\tableaustep\kern\tableaustep\phantomvrule\tableaustep}%
                \hbox{\vbox{\phantomhrule\tableauside}\kern-\tableaurule}}}
\def\squares#1{\hbox{\count0=#1\noindent\loop\sqr
                \advance\count0 by-1 \ifnum\count0>0\repeat}}
\def\tableau#1{\vcenter{\offinterlineskip
                \tableaustep=\tableauside\advance\tableaustep by-\tableaurule
                \kern\normallineskip\hbox
                {\kern\normallineskip\vbox
                        {\gettableau#1 0 }%
                        \kern\normallineskip\kern\tableaurule}%
                \kern\normallineskip\kern\tableaurule}}
\def\gettableau#1 {\ifnum#1=0\let\next=\null\else
        \squares{#1}\let\next=\gettableau\fi\next}

\newcommand{\be}{ \begin{equation}}
\newcommand{\ee}{\end{equation}}
\newcommand{\bea}[1]{\begin{eqnarray}\label{#1} }
\newcommand{\eea}{\end{eqnarray}}

\def\ZZZ{{\hskip-3pt\hbox{ Z\kern-1.6mm Z}}}
\def\zzz{{\hskip-3pt\hbox{ z\kern-1mm z}}}

\def\bal#1\eal{\begin{align}#1\end{align}}

\def\one{{\hbox{ 1\kern-.8mm l}}}
\def\zero{{\hbox{ 0\kern-1.5mm 0}}}

\title{Tensor network and ($p$-adic) AdS/CFT}

\author{Arpan Bhattacharyya${}^{a}$,  Ling-Yan Hung$^{a,b,c}$,  Yang Lei$^{d}$, and Wei Li$^{d}$}

\affiliation{$^a$ Department of Physics and Center for Field Theory and Particle Physics, Fudan University, \\
        \hspace*{0.3cm}220 Handan Road, 200433 Shanghai, P. R. China}

\affiliation{$^b$ State Key Laboratory of Surface Physics and Department of Physics, Fudan University,\\
        \hspace*{0.3cm}220 Handan Road, 200433 Shanghai, P. R. China}
        
\affiliation{$^c$ Collaborative Innovation Center of Advanced  Microstructures,
Nanjing University,\\
        \hspace*{0.3cm}Nanjing, 210093, P. R. China.}
        
\affiliation{$^d$ CAS Key Laboratory of Theoretical Physics, 
Institute of Theoretical Physics,\\ 
\hspace*{0.3cm}Chinese Academy of Sciences, 100190 Beijing, P.R.\ China}

\emailAdd{bhattacharyya.arpan@yahoo.com, elektron.janethung@gmail.com, 
yanglei@itp.ac.cn, 
weili@itp.ac.cn}

\abstract{ 
We use the tensor network living on the Bruhat-Tits tree to give a concrete realization of the recently proposed $p$-adic AdS/CFT correspondence (a holographic duality based on the $p$-adic number field $\mathbb{Q}_p$).
Instead of assuming the $p$-adic AdS/CFT correspondence, we show how important features of AdS/CFT such as the bulk operator reconstruction and the holographic computation of boundary correlators are automatically implemented in this tensor network. 
}

\begin{document}

\maketitle

\makeatletter
\g@addto@macro\bfseries{\boldmath}
\makeatother

\section{Introduction}

The principle of holographic duality states that quantum gravity in a spacetime $\mathcal{M}$ is \textit{equivalent to} a quantum field theory on the boundary $\partial \mathcal{M}$ \cite{'tHooft:1993gx,Susskind:1994vu}.
Both conceptually and mathematically, it provides one of the most powerful tools to understand non-perturbative quantum gravity (via its dual quantum field theory description). 
The most prominent and tractable holographic duality arises when the spacetime is anti-de Sitter (AdS), where we have the AdS/CFT correspondence \cite{Maldacena:1997re}: quantum gravity in an asymptotically AdS  spacetime is equivalent to a conformal field theory (CFT) living on the boundary of the AdS spacetime. 
\smallskip

While the duality was first engineered from string theory,  the Area Law of black holes and general arguments based on the entropy bound suggest that the holographic principle is independent from string theory \cite{'tHooft:1993gx,Susskind:1994vu}. 
How to understand holographic dualities independent of the string theory framework? 
What is the underlying mechanism of AdS/CFT? Proving AdS/CFT would be very difficult because (in most of its parameter regime) it is a strong/weak duality.  
However, short of proving AdS/CFT, is it at least possible to see it emerge from some constructions (other than string theory), i.e.\ without inputing it as an assumption?
\smallskip

Above is the background motivation of the current work. 
Based on the holographic entropy formula by Ryu and Takayanagi, it was noticed by Swingle that certain types of tensor networks can have features of a holographic correspondence \cite{Swingle:2012wq}. 
The observation stimulated a surge of investigations \cite{Czech4, Czech2015, Mintun2015, Nozaki:2012zj,Bao2015,Miyaji1609,Mollabashi,Miyaji201503,references2, references3, Pastawski:2015qua, Hayden:2016cfa, Bhattacharyya:2016hbx,Han:2016xmb,Chirco:2017vhs,Lee:2015vla,Singh:2016mxd,Heydeman:2016ldy,Miyaji201506,Czech1612,Peach1702} on various other types of tensor networks and there is increasing evidence that tensor networks can capture essential features of AdS/CFT. 
\smallskip

An important recent development is the proposal of the holographic code based on perfect tensors on a network embedded in a negatively-curved space \cite{Pastawski:2015qua},\footnote{Perfect tensors emerge very naturally in spaces of tensors with large bond dimensions \cite{Hayden:2016cfa}.} 
which recovers  the RT formula naturally and exhibits the causality that mimics an error correcting code. 
Perturbing away from perfect tensors, other features of AdS/CFT, such as structures similar to Witten diagrams in the computation of correlation functions, also emerge \cite{Bhattacharyya:2016hbx}. 
\smallskip

To further proceed, the study of holographic tensor networks needs to address some conceptual questions. 
A tensor network lives on a discrete space. 
To which extent  can it capture the AdS/CFT correspondence, which has only been defined for continuous spacetimes? 
Does there exist a discrete version of the AdS/CFT correspondence which tensor networks can capture fully? 
\medskip

To answer these questions, in this paper we try to realize the following two important aspects of holography in tensor networks: 
(1) reconstruction of bulk operators  and (2) holographic computation of boundary correlators.
\smallskip

There are two main differences from previous studies of holographic tensor networks.
First of all, we use rather generic tensors. 
The restriction to perfect tensors  in \cite{Pastawski:2015qua} makes it easy to do explicit computations in tensor networks.
However it was later found that in order to have non-trivial correlation functions we need to use imperfect tensors \cite{Bhattacharyya:2016hbx}. 
\smallskip

Second, we propose to put the tensor network on the 
Bruhat-Tits (BT) tree, which is a geometrical presentation of the $p$-adic expansion of a $p$-adic number.
Thus far, the lack of symmetry  has been a bottleneck in the development of tensor networks.\footnote{See however  \cite{Bhattacharyya:2016hbx} that initiated the study on symmetries in the tensor network.}
The main problem is that if we assume that the tensor network realizes a ``naive" discretization of the AdS space, i.e.\ that the tensor network lives on the (dual graph) of a regular tiling of AdS, then the presence of the lattice breaks the continuous isometry group (e.g.\ SL$(2,\mathbb{R})$ in a 2D bulk) down to a \textit{discrete subgroup} of the isometry of AdS. The discrete subgroups preserved by these regular tilings are Coxeter groups, whose representation theory, to our knowledge, is not yet strong enough to give a good description of eigenfunctions on the graph.
\smallskip

To gain more symmetry and quantitative control of the graph wavefunctions, we instead look at tensor networks living on the Bruhat-Tits tree.
The Bruhat-Tits tree preserves  the \textit{full conformal group} SL$(2,\mathbb{Q}_p)$ --- only with the real field $\mathbb{R}$ replaced by a different field completion $\mathbb{Q}_p$ of the rational numbers.
This is a continuous group, and hence much larger than the discrete subgroup of SL$(2,\mathbb{R})$ preserved by any regular tessellation of the AdS\ space. 
Therefore a tensor network based on the Bruhat-Tits tree would have much more symmetry than its counterpart living on a regular tessellation. 
\smallskip

This is inspired by recent proposals for the $p$-adic AdS/CFT correspondence \cite{Gubser:2016guj, Heydeman:2016ldy, Gubser:2016htz}, which generalize the AdS/CFT dictionary to the situation where the boundary theory lives on a space-time that  is based on the  field $\mathbb{Q}_p$  (which is continuous), and where the discrete BT tree plays the role of the bulk AdS space.\footnote{The discussion is mainly based on a two-dimensional bulk and a one-dimensional boundary. The proposed higher-dimensional generalization involves finite algebraic extensions of $\mathbb{Q}_p$, for more details see \cite{Gubser:2016guj, Heydeman:2016ldy}.} 
Besides being an interesting holographic duality that is based on $\mathbb{Q}_p$ instead of $\mathbb{R}$, $p$-adic AdS/CFT might also become relevant to the analysis of the continuous version via the adelic construction.
\smallskip

In this paper we propose to use tensor networks to give concrete realizations of $p$-adic AdS/CFT, in some sense analogous to using various D-brane configurations to engineer corresponding AdS/CFT dualities explicitly.  
This is in contrast to \cite{Gubser:2016guj, Heydeman:2016ldy}, which assumed a $p$-adic AdS/CFT correspondence and then derived various consequences.
In our tensor network construction, we do not \textit{assume} $p$-adic AdS/CFT, but aspects of the AdS/CFT correspondence emerge from the tensor network. 
\smallskip

A possible connection between $p$-adic AdS/CFT and tensor networks was first studied in \cite{Heydeman:2016ldy}, whose construction is based on perfect tensors and on embedding the Bruhat-Tits tree in the geometric tiling (in particular, the HAPPY tiling) of the bulk. 
The construction in the current paper is different in that (1) we are not using perfect tensors, building on earlier work that imperfect tensors are necessary to furnish non-trivial correlation functions \cite{Bhattacharyya:2016hbx}; and (2) we view the Bruhat-Tits tree as an abstract tree, and hence do not need to embed  the Bruhat-Tits tree in a real bulk space. 
Namely, the bulk in $p$-adic AdS/CFT is just the Bruhat-Tits tree itself, and the relation to the real AdS/CFT will not come about by a naive geometric embedding. 
\medskip

The paper is organized as follows. 
In Section \ref{sec:revnet} we review the basics of tensor networks, in particular the intuition behind their role as a discrete holographic correspondence. 
In Section \ref{sec:bulkrec} we generalize the operator-pushing technique developed for perfect tensors in \cite{Pastawski:2015qua} to generic tensors, and then use it to derive a tensor network analogue of the bulk reconstruction formula. 

For generic tensor networks, the results of Section \ref{sec:bulkrec} lack conceptual power because to relate to AdS/CFT we need a notion of conformal or at least scaling primaries. 
Hence in Section \ref{sec:iso} we motivate our proposal of studying tensor networks living on the Bruhat-Tits tree, and show that this allows us to define conformal primaries on tensor networks. 

Building on this, in Section \ref{sec:padicbulkrec} we show that the bulk reconstruction for tree tensor networks gives a nice $p$-adic bulk reconstruction (i.e.\ HKLL) formula. 
We also show a strong parallel between the real and $p$-adic HKLL formulae and in particular that they can both be understood in the linear order as wavelet transforms. 
Section~\ref{correlationfn} computes $p$-adic correlations functions via tensor networks and shows how Witten diagrams emerge in the bulk of the tensor network. 
Finally in Section \ref{sec:summary} we summarize and discuss open questions.  

We leave some review and detailed computation to five Appendices.
Appendix A reviews the lattice construction of the Bruhat-Tits tree and Appendix B reviews the basics of $p$-adic analysis. 
In Appendix C we give two explicit examples. 
Appendix D contains the proof of an argument for the necessity of going beyond perfect tensors. 
Appendix E explains how to realize spacetime symmetry via tensor network transformations. 

\section{Short review of tensor networks} \label{sec:revnet}

In this section we first review basic aspects of tensor networks to fix notations.\footnote{For a good review on tensor networks see \cite{Orus1, Orus2}.} We then explain the intuition behind attempts at using them to realize discrete versions of the holographic duality \cite{Pastawski:2015qua,Hayden:2016cfa}. 
Finally we discuss open questions in tensor networks that motivated this work.

\subsection{Tensor network as ansatz for $N$-body wavefunction}

Solving for exact wavefunctions $\vert \psi \rangle$ of a quantum many-body system analytically is in general a very difficult problem, because of the gigantic dimension of the Hilbert space.
Consider a $N$-body Hamiltonian $\hat{H}$.  
Generically, its wavefunction $\vert \psi \rangle$ is given by a rank-$N$ tensor:
\begin{equation}
\vert \psi \rangle = \sum_{i_1\cdots i_N} f_{i_1\cdots i_N}|i_1\cdots i_N\rangle
\end{equation}
for $1\leq i_k \leq D$, where $D$ is the dimension of the Hilbert space at each site.\footnote{We have assumed that the full Hilbert space of the $N$-body system can be factorized into direct products of Hilbert spaces on each site $\mathcal{H}=\otimes^N_{i=1} \mathcal{H}_i$.}  
Determining the wavefunction $\vert \psi \rangle$  would therefore involve solving  for $D^N$ numbers of unknowns, which scales \textit{exponentially} with $N$. 
\smallskip

The tensor network was introduced as a ``clever" ansatz for $\vert \psi \rangle$ that can greatly simplify the above problem. 
In this ansatz, the rank-$N$ tensor $f_{i_1\cdots i_N}$ in the original wavefunction $\vert \psi \rangle$ is decomposed into many much smaller tensors $T(v)$ (with rank-$r_v$) contracted together: 
\begin{equation} \label{eq:contracttensor}
f_{i_1\cdots i_N}= \sum_{\mu_1,\mu_2 \cdots}T_{i_1\cdots; \mu_1 \mu_2\cdots } (1) T_{i_k\cdots; \mu_1 \mu_3 \cdots}  (2)\cdots.
\end{equation}
In the r.h.s. the original indices from  $i_{1}$ to $i_{N}$ remain un-contracted and we will call them physical or external indices, and $\mu_{n} $ denotes internal indices that are contracted between tensors.  
\smallskip

The contraction of internal indices in the ansatz (\ref{eq:contracttensor}) can be better presented graphically --- in terms of a connected graph $\mathcal{G}$ (or ``network"), where each tensor  $T(v)$ (with rank-$r_v$) is represented by a vertex $v$ with valency $r_v$, each contracted index $\mu_n$ by an edge between two vertices, and each physical index $i_k$ by an external leg at the boundary of $\mathcal{G}$.
\smallskip

One can immediately see that the ansatz (\ref{eq:contracttensor}) can greatly simplify the problem by counting the number of degrees of freedom in the r.h.s.\ of (\ref{eq:contracttensor}).
For a network consisting of $M$ vertices, and for simplicity assuming that they all have the same valency $r$,  the number of total degrees of freedom is then $M D^{r}$. 
Since in a typical tensor network, $M \sim \mathcal{O}(N^d)$ --- here $d$ is a small number that depends on the spacetime dimension and the choice of network --- whereas $r\sim \mathcal{O}(1)$, the tensor network ansatz has far fewer degrees of freedom than the original $N$-body problem:
\begin{equation}
D^N \gg M D^r .
\end{equation}
One can then numerically solve for the ground state wavefunction $|\psi\rangle$ by  minimizing the energy $ \langle \psi| \hat{H} |\psi \rangle$. 
\smallskip

\begin{figure}[h!]
        \centering
        \includegraphics[width=0.7\textwidth]{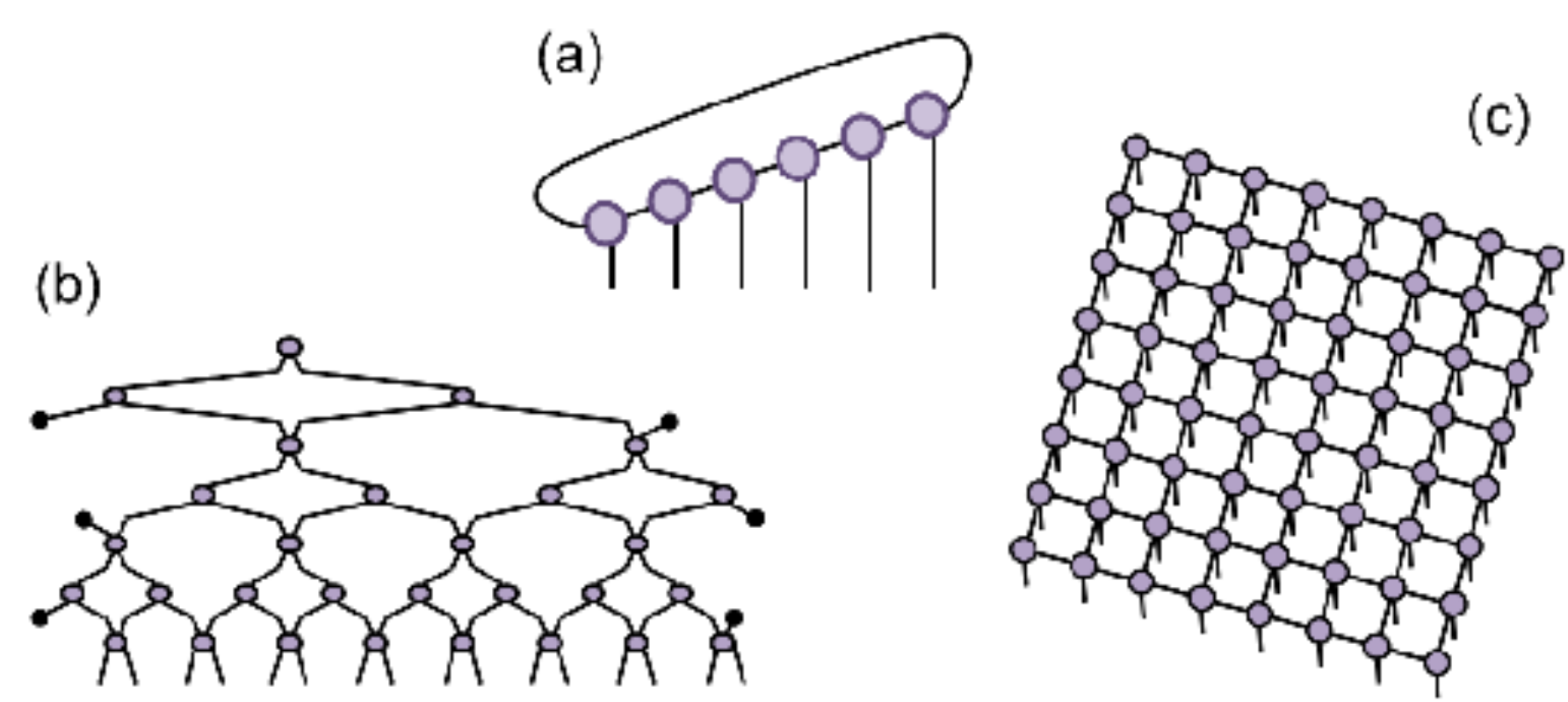}
        \caption{Three well-studied tensor networks. (a) Matrix Product State for (1+1)-dim gapped systems. (b) MERA network for in (1+1)-dim gapless systems. (c) Projected Entangled Pair States  for (2+1)-dim systems. Picture courtesy of \cite{Orus1}. }
        \label{tensornetworks}
\end{figure}

What makes a particular tensor network ansatz (\ref{eq:contracttensor}) numerically efficient and yet remains a good approximation is that the network $\mathcal{G}$ needs to be chosen according to the quantum entanglement structure of the given $N$-body system.
Figure [\ref{tensornetworks}] shows three well-studied tensor network ans\"atze, according to three different types of entanglement structures. 

\subsection{Tensor network as discrete holographic correspondence}

What makes tensor network more than a good computational tool in many-body systems is the realization that certain types of tensor networks can have features of a holographic correspondence. 
This was realized by Swingle in 2011 \cite{Swingle:2012wq}, for the case of 
the MERA (multi-scale entanglement renormalization ansatz) network.
\smallskip

In the MERA network shown in Figure  \ref{tensornetworks}-(b), the physical legs are at the bottom of the network. 
Moving upwards in the network are alternating layers of 4-valent vertices (disentanglers) and those with 3-valent vertices (isometries).\footnote{Without disentanglers, the MERA network reduces to a tree, which (in this naive setting) cannot reproduce the entanglement structure of a gapless system that the MERA network wants to describe.} 
Each set of these twin layers serves as a linear map that projects the system to a coarse-grained one (hence the ``renormalization" in the name). 
As one moves up the graph, a new scale, i.e.\ an extra dimension, emerges, which is very similar to AdS/CFT in which the radial direction of the AdS bulk plays the role of the RG scale.
\smallskip

Moreover, the entanglement entropy of the MERA network is \textit{bounded from above} \cite{Swingle:2012wq} by the length of the geodesic cutting through the network, reminiscent  of the Ryu-Takayanagi formula that computes the entanglement entropy via holography \cite{Ryu,Ryu1}. 

\subsubsection{Bulk and boundary in tensor networks}

To establish a more concrete connection between tensor networks and holography, let's first define the meaning of bulk and boundary in the context of tensor networks.
The network $\mathcal{G}$ plays the role of the (discrete) bulk space. 
The bulk Hilbert space is defined as follows. 
First, assign a  $D$-dim Hilbert space $\mathcal{H}_e$ to each edge,\footnote{We assume that the bond dimension $D$ for each edge, i.e.\ the size of each $\mathcal{H}_{e}$, is the same.} and a $D^r$-dim Hilbert space $\mathcal{H}_v \equiv \otimes_{e(v)=1}^r \mathcal{H}_{e(v)}$ to each vertex $v$. 
(Here $e(v)$ denotes an edge emitting from vertex $v$.)  
The tensor $T(v)$ then essentially defines a state in $\mathcal{H}_v$
\begin{equation} \label{eq:vertex_state}
|T(v)\rangle \equiv \sum_{\{a_i\}}
T_{a_1\cdots a_r} (v)|a_1 \cdots a_r \rangle.
\end{equation}
The bulk Hilbert space is then
\begin{equation}
\mathcal{H}_{\textrm{bulk}} \equiv \otimes^M_{v=1} \mathcal{H}_v ,
\end{equation}
where $M$ is the total number of vertices in the network.
A generic bulk state is then
\begin{equation}
|\Psi_{\textrm{bulk}}\rangle = \sum_{\{T(v)\}}\alpha_{\{T(v)\}}  \left( \otimes_{v} |T(v) \rangle \right).
\end{equation}
In the simplest case (as in section 2.1), it is a product state $|\Psi_{\textrm{bulk}}\rangle = \otimes_{v} |T(v) \rangle$.
\smallskip

The boundary Hilbert space is just the original $\mathcal{H}$ defined at the beginning of this section:
\begin{equation}
\mathcal{H}_{\textrm{bndy}} = \otimes^{N}_{i=1} \mathcal{H}_{i},
\end{equation}
where $\mathcal{H}_{i}$ is the Hilbert space living on the external leg $i$ of the network $\mathcal{G}$.
\smallskip

The map from bulk to boundary is via a projection operator that effectively contracts all the internal indices: 
\begin{equation} \label{eq:holographicmap}
 |\psi_{\textrm{bndy}}\rangle = \left( \otimes_{\textrm{all internal edges $\mu$}} | \mu  \rangle \langle \mu|\right) \,  |\Psi_{\textrm{bulk}}\rangle.
\end{equation}
For each internal edge $\mu$ (connecting two vertices $v$ and $w$), $| \mu \rangle$ is defined as 
\begin{equation}
| \mu \rangle \equiv  | \mu_{v,w} \rangle = \sum_{\alpha_{\mu_{v}}, \alpha_{\nu_{w}}}\kappa_{\alpha_{\mu_{v}} \alpha_{\nu_{w}}}  |\alpha_{\mu_{v}} \rangle \otimes | \alpha_{\nu_{w}} \rangle ,
\end{equation}
where $|\alpha_{\mu_{v}}\rangle $ denotes the state located at the edge-$\mu$ of the vertex $v$. 
The boundary wavefunction is thus defined by projecting the edges connecting two vertices into an entangled state.   
In most cases considered, such as our example described in (\ref{eq:contracttensor}) and in the rest of the paper, the ``metric"   $\kappa_{\alpha_{\mu_{v}} \alpha_{\nu_{w}}} = \delta_{\alpha_{\mu_{v}} \alpha_{\nu_{w}}} $.

\subsubsection{Perfect tensor code and operator pushing}

An important shortcoming of the MERA network as discrete holography is  that the Ryu-Takayanagi formula does not compute its entanglement entropy but only provides an upper bound. 
Part of the reason is that its network does not preserve enough  of the hyperbolic isometries expected for the dual theory of a CFT.

\smallskip
The ``HAPPY" code in \cite{Pastawski:2015qua} improved this by choosing the network $\mathcal{G}$ to be the dual graph of a regular tessellation of the hyperbolic space. 
With the further assumption that the tensors are restricted to be ``Perfect Tensors'', the HAPPY code was the first tensor network that exactly recovers the Ryu-Takayanagi formula \cite{Pastawski:2015qua}.

\smallskip
The restriction to ``Perfect tensors" greatly simplifies the derivation in \cite{Pastawski:2015qua}.  
They are  even-rank tensors with the following property: for any partition of the $r$ indices into two sets $n \in\{\mu_1,\dots \mu_{k}\}$ and $a\in\{\mu_{k+1},\dots \mu_{r}\}$ with $k \leq \frac{r}{2}$, $T_{na}$ is a norm-preserving projection operator from $a$ to $n$:
\begin{equation}\label{eq:perfectdef}
T^{\phantom{\dagger}}_{na} T_{a n'}^{\dagger}=D^{\frac{r}{2}-k} \, \delta_{nn'} .
\end{equation}
In particular, when $k= \frac{r}{2}$,  $T_{na }$ becomes a unitary map: $T^{\phantom{\dagger}}_{na}T_{a n'}^{\dagger}=\delta_{nn'}$.
\smallskip

Another important result in \cite{Pastawski:2015qua} is the exhibition of the causal structure in the HAPPY code:  
it was shown that a operator acting on the bulk of the HAPPY code can be recovered using only boundary operators that act on a subregion of the boundary.\footnote{This is analogous to the bulk operator in the Rindler coordinates, in which only the subregion within the causal wedge containing  the original bulk operator is involved.}
\smallskip

This was shown using the method of ``operator pushing", invented for the perfect tensor by \cite{Pastawski:2015qua}.
Let's illustrate this method with a particular example of a perfect tensor code: the hexagon code (i.e. $r$=6) with bond dimension $D=2$. 
\smallskip

A tensor state $ |T\rangle$ in the hexagon code is invariant under a set of $6$ stabilizers $S^{(a)}$:
\begin{equation}\label{Stable} 
S^{(a)} |T\rangle^{\textrm{hexagon}}  = |T\rangle^{\textrm{hexagon}} \qquad \qquad a =1,2,\dots,6.\quad 
\end{equation}
Since $D=2$, $S^{(a)}$ can be expressed in terms of Pauli matrices $\{X,Y,Z\}$ acting on the 6 edges. 
We can choose a basis such that\footnote{For the full list of stabilizers see \cite{Pastawski:2015qua}.}
\begin{equation}\label{S1}
S^{(1)} = X_1\otimes X_2 \otimes X_3\otimes X_4 \otimes X_5 \otimes X_6 \qquad  \textrm{etc}
\end{equation}
where $X_i$ acts on the $i^{\textrm{th}}$ leg of $ |T\rangle$. 
We immediately see that (\ref{Stable}) with the stabilizer (\ref{S1}) implies 
\begin{equation}\label{OPhex}
\forall v: \qquad X_i |T\rangle^{\textrm{hexagon}}  = \otimes_{j\neq i} \, X_j|T\rangle^{\textrm{hexagon}} \end{equation}
Namely, applying the operator $X$ one of the legs of $|T\rangle$ is equivalent to applying $X$ on the other five legs. 
\smallskip 

Applying other stabilizers gives similar ``operator pushing" rules, in which an operator acting on a given leg of a bulk tensor site can be ``pushed" to the other legs of the same site. 
Applying the set of rules repeatedly, one can move the effect of a bulk operator all the way to the boundary, using which \cite{Pastawski:2015qua} showed  the emergence of the causal structure in the perfect tensor code. 
The main idea can be generalized to general tensors and will be used later to derive the tensor network analogue of the HKLL formula.

\subsection{Some open questions in tensor network as discrete AdS/CFT} \label{sec:openq}

For tensor networks to realize a discrete version of AdS/CFT, the following aspects need to be further developed. 

\begin{enumerate}

\item Global symmetries of the boundary wavefunction v.s. isometries of the network.

In traditional AdS/CFT, gauge symmetries of the bulk are mapped to global symmetries of the boundary theory. 
In particular, the isometries of the bulk spacetime are mapped to global spacetime symmetries of the boundary CFT. 
How to realize the interplay between bulk isometries and boundary global symmetries in the tensor network? 
Further, how to effectively use the symmetry in the tensor network computation to model discrete AdS?\footnote{Some discussion on how tensor networks reproduce such features already appeared in \cite{Guifre1}.}

\item Going beyond perfect tensors

Though greatly simplifying the computation, perfect tensors are in some sense ``too perfect" --- namely too symmetric to  allow enough dynamical content. 
For example, in a tensor network with only perfect tensors there exists no connected correlation function between local operators \cite{Bhattacharyya:2016hbx}.

\item Tensor network analogue of the HKLL formula 

As we demonstrate in Appendix E, in a tensor network using perfect tensors, a bulk operator can be reconstructed only as macroscopic products of boundary operators. 
This is in sharp contrast with the HKLL formula where in the large $N$ limit, the leading contribution is linear in the boundary operators. 
We need to generalize the operator-pushing technique invented for perfect tensors to more generic tensors, and find a 
tensor network analogue of the HKLL formula.

\end{enumerate}

In this paper, we will take some further steps in clarifying these questions.

\section{Bulk operator reconstruction}
\label{sec:bulkrec}
 
In this section we show how the bulk operator reconstruction is realized tensor networks. 
This was initiated for the special case of perfect tensors in \cite{Pastawski:2015qua}. 
We first show that in order to have the analogue of the HKLL formula one needs to use more general tensors, then derive an analogue of the HKLL formula for tensor networks.

\subsection{From HKLL to operator pushing}

In AdS/CFT, a normalizable local field in the bulk $\phi^I(\vec{x}, z)$ can be reconstructed from the boundary operators via the HKLL formula \cite{Hamilton:2006,Hamilton:2005ju}\footnote{Here $(\vec{x},z)$ denotes the  Poincar\'e coordinate with the metric $ds^2=\frac{1}{z^2}(dz^2+d\vec{x}^2)$ and $\vec{x}$ is the boundary $d$-vector with $\vec{x}^2\equiv \eta_{\mu\nu}x^{\mu}x^{\nu}$ with negative signature.} 
\begin{equation}\label{fullHKLL}
\begin{aligned}
&\phi^I(\vec{x}, z )=\int d^dy  \, K_I(\,\vec{x},z \,\vert \vec{y}\,)\, \mathcal{O}^I (\vec{y})\\
&+\sum_{J,K}\frac{\lambda^I_{JK}}{N}\int d^dx' dz' \,G_I(\vec{x},z\vert \vec{x}',z')\int d^dy_1\,K_J(\vec{x}',z' \vert \vec{y}_1)\mathcal{O}^J (\vec{y}_1) \int d^dy_2 \,  K_K(\vec{x}',z' \vert \vec{y}_2) \mathcal{O}^K (\vec{y}_2)\\
&+\mathcal{O}(\frac{\lambda^2}{N^2})\int\dots
\end{aligned}
\end{equation}
where $\mathcal{O}^I$ is the boundary operator that is dual to the bulk field $\phi^I$: $\lim_{z\rightarrow 0} \phi^I(\vec{x},z) = z^{-\Delta_I}  \mathcal{O}^I(\vec{x})$ (with $\Delta_I$ the conformal dimension of $\mathcal{O}^I$).
Here $K_{I}(\,\vec{x},z \,\vert \vec{y}\,)$ is the boundary-bulk kernel (called ``smearing function" here) of $\mathcal{O}^I$:
\begin{equation}\label{Kkernel}
K_I(\vec{x},z \, \vert \, \vec{y}\,)= \left(\frac{z}{z^2-(\vec{x}-\vec{y})^2 }\right)^{d-\Delta_I} \Theta(z^2-(\vec{x}-\vec{y})^2)
\end{equation}
and $G_I(\vec{x},z\vert \vec{x}',z')$ its bulk-bulk kernel. 
The ``smearing function"  and the bulk-bulk kernel are simply related by 
$\lim_{z'\rightarrow 0}z'^{\Delta_I-d}G^I(\vec{x},z\vert \vec{x}',z')\sim 
K^I(\vec{x},z\, \vert\, \vec{x}'\,)$. 
Note that the bulk-bulk kernel here satisfies a different boundary condition from the usual bulk propagator.
\smallskip

For a tensor network to be a (discrete version) of the holographic duality (\ref{fullHKLL}), it needs to realize an analogue of the HKLL formula, i.e.\ an operator acting on the $|\Psi_{\textrm{bulk}}\rangle$ should be reconstructed by operators acting on the $|\psi_{\textrm{bndy}}\rangle$, in a way similar to (\ref{fullHKLL}). 
In particular, it should exhibit the causal structure (i.e.\ the reconstruction should be possible using only a subregion)  
and an expansion in which the linear term dominates in the large-$N$ limit.
\smallskip

The emergence of the causal structure in tensor networks was realized by \cite{Pastawski:2015qua} for a specific tensor network with ``perfect tensors". 
Motivated by the observation by \cite{references2} that the causal structure in the HKLL formula (\ref{fullHKLL}) resembles the error correction code, \cite{Pastawski:2015qua} showed that a bulk operator in the perfect code can be recovered using only those boundary operators that live on the subregion that is causally connected to the original bulk operator. 
\smallskip

However, in a network with perfect tensors, a bulk operator can be reconstructed only as a macroscopic product of boundary operators. 
This is in sharp contrast with the HKLL formula in which the leading term in the large-$N$ limit is linear in the boundary operators. 
Thus a tensor network based on ``perfect tensors" does not lead to an analogue of the HKLL formula. 

\subsection{Operator pushing}

In section 2.2 we briefly reviewed the idea of operator pushing for perfect tensors: 
using the stabilizer relations one can ``push" the operator acting on a given leg of a tensor to pass this tensor and become operators acting on the other legs of the same tensor \cite{Pastawski:2015qua}. 
Now we generalize this method for more general tensors, in order to derive the tensor network analogue of the HKLL formula. 

\subsubsection{Local operator pushing}

Given a vertex state $|T(v)\rangle$  defined in (\ref{eq:vertex_state}) --- recall that none of its vertices are contracted, and hence there is no difference between internal and external legs yet --- we apply the operator $P^A$ on its $a^{\textrm{th}}$ leg. 
Using the non-degenerateness of $T$ (see below), the result can always be rewritten as
\begin{equation}\label{eq:push1}
\begin{aligned}
P^A(a) |T(v)\rangle&= \sum_{b\neq a} \sum_{B} \, \alpha^{A}_{B}(a|b)\,P^B(b)\,|T(v)\rangle\\
&+\sum_{b\neq c \neq a} \sum_{B,C} \, \alpha^{A}_{BC}(a|b,c)\,P^B(b)\, P^C(c)\, |T(v)\rangle \\
&+\sum_{b\neq c \neq d  \neq a} \sum_{B,C,D} \, \alpha^{A}_{BCD}(a|b,c,d)\, P^B(b)\, P^C(c)\, P^D(d)\, |T(v)\rangle +\dots.\\
\end{aligned}
\end{equation}
Here $P^A(a)$ denotes the operator $P^A$ acting on the state living on the $a^{\textrm{th}}$ leg of the vertex $v$, the sum of $b,c,\dots$ are over the $r_v$ legs of vertex $v$, and the sum of $B,C,\dots$ is over the operator basis excluding the identity operator. 
Generically, $\alpha$ depends on the value of the tensor state $ |T(v)\rangle$.
In the simple example of the hexagon code (\ref{OPhex}), $\alpha^{X}_{XXXXX}(1|23456)=1$ and similarly for the other legs.\footnote{In general these coefficients $\alpha$ are not unique. 
In the case of perfect tensors, they are defined up to actions of the list of stabilizers.} 
\smallskip 

We are using the convention that an identity operator acting on an edge is the same as nothing acting on it. 
For example, in $\alpha^{I}_{J_1 J_2 \dots J_{r-1}}$ where $r$ is the rank, if say operators from $J_2$ to $J_{r-1}$ are all identity operators, we then write it as $\alpha^{I}_{J_1}$. 
This is to distinguish between different orders of operator-pushing coefficients, i.e.\ the number of non-identity operators on outgoing legs.\footnote{Sometimes it is convenient to consider all operator pushing coefficients on an equal footing, then we sum over the operator basis including the identity operator and explicitly keep the form $\alpha^{I}_{J_1 J_2 \dots J_{r-1}}$ even if some operators in $\{J_k\}$ are identity operators} 
\smallskip

\begin{figure}[h!]
        \centering
        \includegraphics[trim=0.5cm 20cm 5cm 3cm, width=0.5\textwidth]{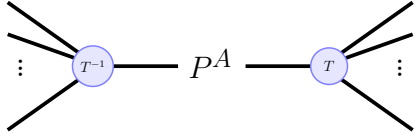}
        \caption{Left hand side of equation (\ref{eq:push}).
        }
        \label{tensorcontract4}
\end{figure}

Next we explain how we can solve for the local operator-pushing coefficients $\alpha$ in (\ref{eq:push1}) systematically. 
Expressing the tensor state $|T(v)\rangle$ explicitly in terms of the rank-$r_v$ tensor (using (\ref{eq:vertex_state})), one can invert the tensor $T$, and sandwich the operator $P^A$ by $T$ and $T^{-1}$. 
Here $T^{-1}$ is considered an inverse in the sense that it satisfies
\begin{equation} \label{eq:weakinv}
\begin{aligned}
& T^{-1}_{a_1  \,a_2\, a_3  \dots a_{r}} T^{\phantom{-1}}_{\tilde{a}_1 \, a_2 \,a_3\dots a_{r}} =\, \delta_{a_1 \,\tilde{a}_1} \end{aligned}
\end{equation}
We can rewrite equation (\ref{eq:push1}) as follows:
\begin{equation} \label{eq:push}
\begin{aligned}
&T^{-1}_{a\, b_1 \cdots b_{r-1}} P^A_{a a'} T_{a' b_1'\cdots b_{r-1}'}= \bigg[\sum_{b\neq a} \sum_{B} \, \alpha^{A}_{B}(a|b)\,P^B(b) \otimes^{r-2} \mathbb{I} \,\\
&+\sum_{b\neq c \neq a} \sum_{B,C} \, \alpha^{A}_{BC}(a|b,c)\,P^B(b)\, P^C(c) \otimes^{r-3} \mathbb{I}\, \\
&+\sum_{b\neq c \neq d  \neq a} \sum_{B,C,D} \, \alpha^{A}_{BCD}(a|b,c,d)\, P^B(b)\, P^C(c)\, P^D(d) \otimes^{r-4} \mathbb{I} \, +\dots \bigg]_{b_1 \cdots b_{r-1}, \,\,  b_1'\cdots b_{r-1}'}\\
\end{aligned}
\end{equation}
This is illustrated in Fig. \ref{tensorcontract4}.
\smallskip

Recall that we assume that the bond dimension $D$ is the same for all edges in the network. 
Thus we can simply use the same basis for the set of $P^{A}$ across the network.
First, we only need to consider traceless operators. 
A convenient choice of basis for the local operator pushing is the set of  $D\times D$ (generalized) Pauli matrices:
\begin{equation} \label{eq:genPaul}
\textrm{Tr}(P^A P^B) = D\, \delta^{AB} \qquad \textrm{with}\quad A, B= 1,2,\dots,D^2-1
\end{equation}
As we will see later, a convenient basis for the global operator pushing will be different from the set $\{P^A\}$.
\smallskip

\begin{figure}[h!]
	\centering
	\includegraphics[trim=2cm 21cm 5cm 4cm, width=0.5\textwidth]{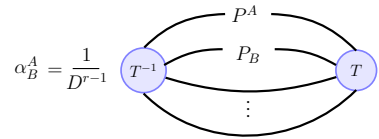}
	\caption{Linear part of operator pushing coefficient \eqref{eq:push1}.}
	\label{fig:alphaab}
\end{figure}

Using the tracelessness of $P^{A,B}$, we have
\begin{equation}\label{alpha2}
\begin{aligned}
& \alpha^A_B (1|2)=\frac{1}{D^{r-1}} \, 
P^{\phantom{A}}_{B, \,a_{2} \, \tilde{a}_2} \, 
 T^{-1}_{ a_1 \, \tilde{a}_2\,a_3 \dots a_{r}}  \,
P^{A}_{\phantom{B}a_1\, \tilde{a}_{1}}\,
T^{\phantom{-1}}_{\tilde{a}_{1}  \, a_{2} \,a_3\dots a_{r} }\,
\end{aligned}
\end{equation}
where we use lower indices to denote the transpose:
\begin{equation}
P_A\equiv(P^A)^t 
\end{equation}
and similarly for index permutations. 
We draw this local ``operator pushing" matrix (\ref{alpha2}) in Figure \ref{fig:alphaab}. 
%%%%%%%%%%%%%%%%%%%%%%%%%%
We have assumed that the tensor $T$ is non-degenerate to the extent that $T^{-1}$ satisfying (\ref{eq:weakinv}) exists. 
Equation (\ref{eq:weakinv}) assumes a normalization whose physical meaning will be apparent as we inspect correlation functions.\footnote{It should be noted that the normalization convention we choose in (\ref{eq:weakinv}) is different from (\ref{eq:perfectdef}) that is used in \cite{Pastawski:2015qua}.}
Eq. (\ref{alpha2}) shows the coefficient corresponding to pushing the operator $P^A$ through the leg-$1$ to the operator $P^B$ acting on leg-$2$. 
Other configurations can be obtained similarly.  
\smallskip 

Similarly for pushing the operator $P^A$ acting on leg-$1$ into the operator   $P^B$ acting on leg-$2$ together with the operator  $P^C$ acting on leg-$3$:
\begin{equation} \label{eq:3ptalpha}
\begin{aligned}
& \alpha^A_{BC} (1|2,3)=\frac{1}{D^{{r-1} }}  \, 
 T^{-1}_{a_1 \, \tilde{a}_2\, \tilde{a}_3\, a_4 \dots a_{r}}  \,
P^{\phantom{A}}_{B, \, a_2\, \tilde{a}_{2}} \, 
P^{\phantom{A}}_{C, \, a_3\, \tilde{a}_{3}} \, 
P^{A}_{\phantom{B}a_1 \tilde{a}_{1}}\,
T^{\phantom{-1}}_{\tilde{a}_{1}  \, a_{2}\,a_{3} \,a_4\dots a_{r} }\,
\end{aligned}
\end{equation}
Higher coefficients can be computed in a similar manner.\footnote{To discuss all operator-pushing coefficients on an equal footing, we can append the identity operator to the list of Pauli matrices $\{P^{I}\}$ with $I=1,\dots,D^2-1$, and define $\mathbf{1}\equiv P^{0}$; then $\alpha^{A}_{B_1, \dots B_{r-1}}$ cover all coefficients.}
\smallskip

An immediate consequence of the above computations is that for perfect tensors (defined in (\ref{eq:perfectdef})), the first non-zero operator pushing coefficient appears at $\alpha^{A}_{B_1 \dots B_{r/2-1}}(a|b_1\dots b_{r/2-1})$.

\subsubsection{Global operator pushing}

Now we can follow the rule of local operator pushing to move an operator acting in the bulk of the network $\mathcal{G}$ all the way to its boundary. 
Consider an operator $P^I$ acting on the $i^{\textrm{th}}$ leg of the vertex $v$ in the bulk of the network. 
Using the map from the bulk wavefunction to the boundary one (\ref{eq:holographicmap}), we get the effect of $P^I$ on the boundary wavefunction:
\begin{equation}\label{bulk0}
\begin{aligned}
 &\left( \otimes_{\textrm{all internal edges $\mu$}} | \mu  \rangle \langle \mu|\right)P^I(v,i) |\Psi_{\textrm{bulk}}\rangle \\
= &\sum^N_{j=1} \sum_{J} \, \mathcal{A}^{I}_{J}(v,i|j)\,P^J(a)\,|\psi_{\textrm{bndy}}\rangle\\
+&\sum_{j\neq k} \sum_{J,K} \, \mathcal{A}^{I}_{JK}(v,i|j,k)\,P^J(j)\, P^K(k)\, |\psi_{\textrm{bndy}}\rangle \\
+&\sum_{j\neq k \neq\ell  } \sum_{J,K,L} \, \mathcal{A}^{I}_{JKL}(v,i|j,k,\ell)\, P^J(j)\, P^K(k)\, P^L(\ell)\,|\psi_{\textrm{bndy}}\rangle +\dots\\
\end{aligned}
\end{equation}
Each coefficient $\mathcal{A}$ is the result of the contributions from every local-operator pushing coefficient $\alpha$ on each leg of the (branched) path from the initial leg (the $i^{\textrm{th}}$ leg of vertex $v$) to the final boundary positions, and then summed over all possible such branched paths though the bulk. 
Below we will explicitly compute $\mathcal{A}^{I}_{J}(v,i|j)$ and $\mathcal{A}^{I}_{JK}(v,i|j,k)$.

\subsection{Linear order in HKLL}

\begin{figure}{}
 \centering
 \includegraphics[trim=6cm 16cm 7.5cm 3cm, width=0.4\textwidth]{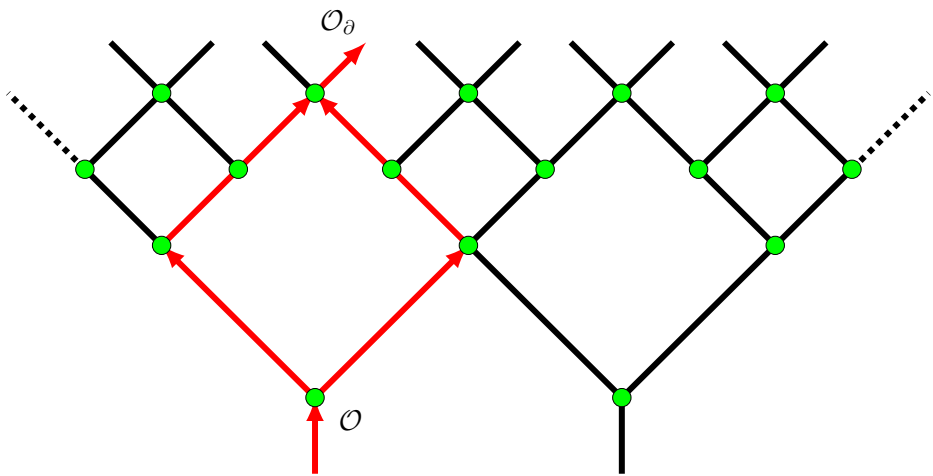}
\caption{Linear operator pushing on a generic tensor network. The red paths indicate all the paths joining the bulk operator and the boundary operator. }
 \label{fig:tnhkll}
\end{figure}

Now we compute the tensor network analogue of the linear term of the HKLL formula (\ref{fullHKLL}), i.e.\ the linear global operator pushing coefficient $\mathcal{A}^{I}_{J}(v,i|j)$.
It is given by the sum over products of $\alpha^{I}_{J}(v_i)$ collected along all the paths $\mathcal{P}$ from the bulk vertex $v$ to the boundary leg $j$
\begin{equation} \label{HKLLemerge}
\mathcal{A}^{I}_{J}(v,i|j) =  \sum_{\mathcal{P}} \sum_{\{I_{1},\dots I_{|\mathcal{P}|-1}\}}\prod_{v_i\in \mathcal{P}}  \alpha^{I_{i-1}}_{I_{i}}(v_i),
\end{equation}
where  $I_{0}\equiv I$, $v_0\equiv v$ and  $I_{|\mathcal{P}|}\equiv J$ (here $|\mathcal{P}|$ is the length of the path $\mathcal{P}$). 
See Figure \ref{fig:tnhkll}.
\smallskip

Now comes an important simplification. 
Since in this paper we are interested in using tensor networks to model AdS/CFT, the bulk network should be homogenous, therefore we can drop all the vertex dependence. 
The equation above simplifies into
\begin{equation} \label{HKLLemergeS}
\mathcal{A}^{I}_{J}(v,i|j) =  \sum_{\mathcal{P}} [(\alpha)^{\otimes |\mathcal{P}(v\rightarrow j)|}]^I_J
\end{equation}
where $\alpha$ denotes the matrix $\alpha^{I}_{J}$ that contains the local operator pushing coefficient from operator $P^I$ to operator $P^J$ (where $\{P^I\}$ is the Pauli matrix basis) and it takes the same value throughout the network. 
\smallskip

We then diagonalize $\alpha$ and use its eigenvectors (labeled as $\mathcal{O}^I$ with eigenvalue $\lambda_{I}$) as the basis for the bulk dual of primary operators.\footnote{Their normalization will be fixed by normalizing the two-point correlation functions later.} 
Using this new basis, the linear term of the global operator-pushing coefficient defined as in (\ref{bulk0}) is simply
\begin{equation} \label{HKLLemergeD}
\mathcal{A}^{I}_{J}(v,a|i) = \delta^{I}_{J} \, K_I(v|i)
\end{equation}
with the ``smearing function" given by
\begin{equation}
K_I(v|i)\equiv \sum_{\mathcal{P}}(\lambda_I)^{|\mathcal{P}(v\rightarrow i)|}
\end{equation}
where the sum is over all paths connecting the vertex $v$ and the boundary leg $i$.
\smallskip

Plugging (\ref{HKLLemergeD}) into the linear term of the bulk operator reconstruction (\ref{bulk0}) from the global operator pushing, we see that it has exactly the same form as the HKLL formula:
\begin{equation}\label{HKLLemergeF}
\mathcal{O}^I(v,1) \, |\Psi_{\textrm{bulk}}\rangle =\sum_{i} K_I (v|i)\mathcal{O}^I(i)\, |\psi_{\textrm{bndy}}\rangle
\end{equation}
Namely, a operator acting $\mathcal{O}^I$ in the bulk can be ``reconstructed", to linear order, by a sum over the same operator acting on the boundary edges weighted by the  ``smearing function" $K_I$.

\subsection{Non-linear orders in HKLL}

Now we move on to the non-linear terms in the global operator pushing formula (\ref{bulk0}).  
Using the same basis that diagonalizes the  linear order global operator pushing in (\ref{HKLLemergeS}), we now compute the coefficient $\mathcal{A}^I_{JK}(v,i|j,k)$, which corresponds to pushing the operator $\mathcal{O}^I$ all the way to the two boundary operators $\mathcal{O}^J$ acting on the boundary leg $j$ and $\mathcal{O}^K$ on the boundary leg $k$.

\begin{figure}
	\centering
	\includegraphics[trim=6cm 16cm 7.5cm 3cm, width=0.4\textwidth]{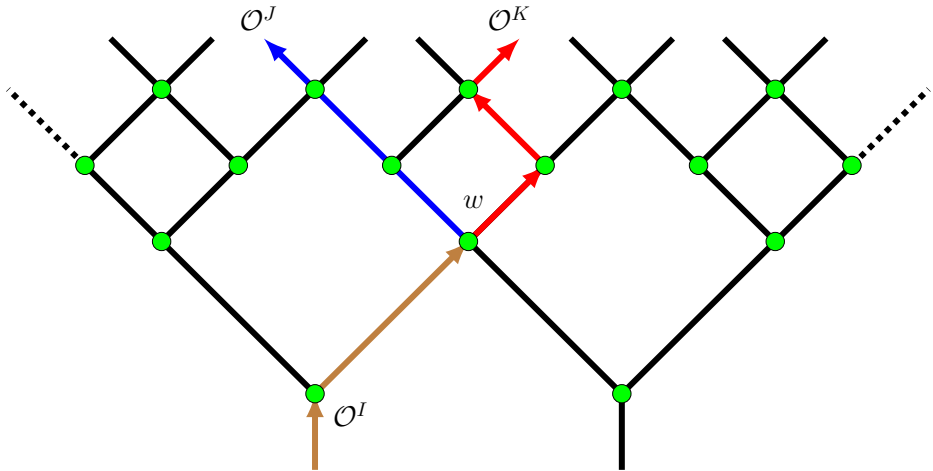}
	\caption{``Non-linear'' contributions to operator pushing. The bulk operator $\mathcal{O}^I$ splits into $\mathcal{O}^J$ and $\mathcal{O}^{K}$. They are subsequently propagated to   $1$ (blue) and $2$ (red) respectively.}
	\label{fig:tnnonlinear}
\end{figure}

\smallskip
Now we use the same argument as the one for the linear term of the global pushing. 
Since there are two operators at the boundary links $j,k$, the contribution to $\mathcal{A}^I_{JK}(v,i|j,k)$ involves splitting the operator $\mathcal{O}^I$ into $\mathcal{O}^J$ and $\mathcal{O}^K$, at some bulk vertex $w$. 
A contribution comes from a product of local operator-pushing coefficient $\alpha^I_I$ along the path $\mathcal{P}$ that joins the initial vertex $v$ to the mid-way bulk vertex $w$. 
At $w$ we use the local operator pushing coefficient $\alpha^I_{JK}$ obtained in (\ref{eq:3ptalpha}) to split the operator $\mathcal{O}^I$ into $\mathcal{O}^J$ and $\mathcal{O}^K$. 
Then we push these operators $\mathcal{O}^J$ and $\mathcal{O}^K$ along paths $\mathcal{P}(w\rightarrow j)$ and $\mathcal{P}(w \rightarrow k)$, respectively.
Finally, we sum over all paths and the mid-way bulk vertex $w$.  (See Figure \ref{fig:tnnonlinear}.)
\smallskip

The final result is 
\begin{equation}\label{nonlinearA}
\mathcal{A}^I_{JK}(v,i|j,k) = \sum_{w}G_I(v|w)\, \alpha^{I}_{JK} \, K_J(w\, |\, j) K_K(w\,|\,k),
\end{equation}
where we have defined the bulk-bulk kernel
\begin{equation}\label{TNbulkG}
G_I(v|w) \equiv \sum_{\mathcal{P}} \lambda_I^{|\mathcal{P}(v\rightarrow w)|}
\end{equation}
where the sum is over all paths that connect two bulk vertices $v$ and $w$.
Note that the bulk-bulk kernel $G_I(v|w)$ and the ``smearing function" $K_I(v|j)$ are simply related by taking the $w$ all the way to the boundary leg $j$.\footnote{For infinite networks, we need to regularize this limit.} 
This can be compared with the bulk-bulk reconstruction kernel in \cite{Kabat:2011rz} that we quoted in (\ref{fullHKLL}). 
\smallskip

Higher order coefficients $\mathcal{A}^I_{JKL\cdots}(v,i|j,k,\ell\cdots)$ can be obtained in the same way. 
For example, 
\begin{equation}
\begin{aligned}
\mathcal{A}^I_{JKL}(v,i|j,k, \ell) & =\sum_{w}\,G_I(v|w)\, \alpha^{I}_{JKL}\, K_J(w|j) \, K_K(w|k) \, K_L(w|\ell)  \\
&+ \sum_{w, u}\, \bigg(G_I(v|w) \,\alpha^I_{JM}\,K_J(w|j) \,G_M(w| u) \, \alpha^{M}_{KL} \, K_K(u|k)\, K_L(u|\ell)  \\
&\qquad + G_I(v|w) \,\alpha^I_{KM}\, K_K(w| k) \,G_M(w|u)\, \alpha^{M}_{JL} \,K_J(u|j)\, K_L(u|\ell)\\
&\qquad + G_I(v|w) \,\alpha^I_{ML}\, K_L(w| \ell) \,G_M(w|u)\, \alpha^{M}_{JK} \,K_J(u|j)\, K_K(u|k) \bigg) .
\end{aligned}
\end{equation}
\smallskip

For a tensor network defined on a tree, this is the complete set of contributions. 
In a generic network with loops, these contributions will be dressed by loop diagrams. 
A complete analysis of these loop diagrams is beyond the scope of the current paper. 
In the following,  we will focus on tree networks. 

\section{Tensor networks on $p$-adic tree}
\label{sec:iso}

In this section we motivate the study of tensor networks living on the $p$-adic tree as an explicit example of discrete AdS/CFT. 
We emphasize its difference from tensor networks based on regular tessellations of AdS.
It can be viewed as an explicit realization of $p$-adic AdS/CFT recently proposed in \cite{Gubser:2016guj, Heydeman:2016ldy}.\footnote{Note that the relation of $p$-adic AdS/CFT to the tensor network was also mentioned in \cite{Heydeman:2016ldy}, although it was based on geometric embedding of the $p$-adic tree in regular tessellation of AdS, as opposed to an abstract tree. }

\subsection{From tessellation to tree}
\subsubsection{Limitation of tensor networks based on regular tessellation}

In the AdS/CFT correspondence, the global symmetries of the boundary CFT are mapped to isometries of the bulk spacetime. 
When using tensor networks to model holography, the network $\mathcal{G}$ corresponds to the bulk space. 
To realize a discrete version of AdS, the most common approach is to choose $\mathcal{G}$ based on a regular tessellation. 
The bulk isometry is then the discrete subgroup of SL$(2,\mathbb{R})$ preserved by the particular tessellation.
\smallskip

\begin{figure}[htb]
\center
\subfigure[][]{
\label{fig:H246}
\includegraphics[trim=0.5cm 5cm 0.5cm 4cm,width=0.3\textwidth]{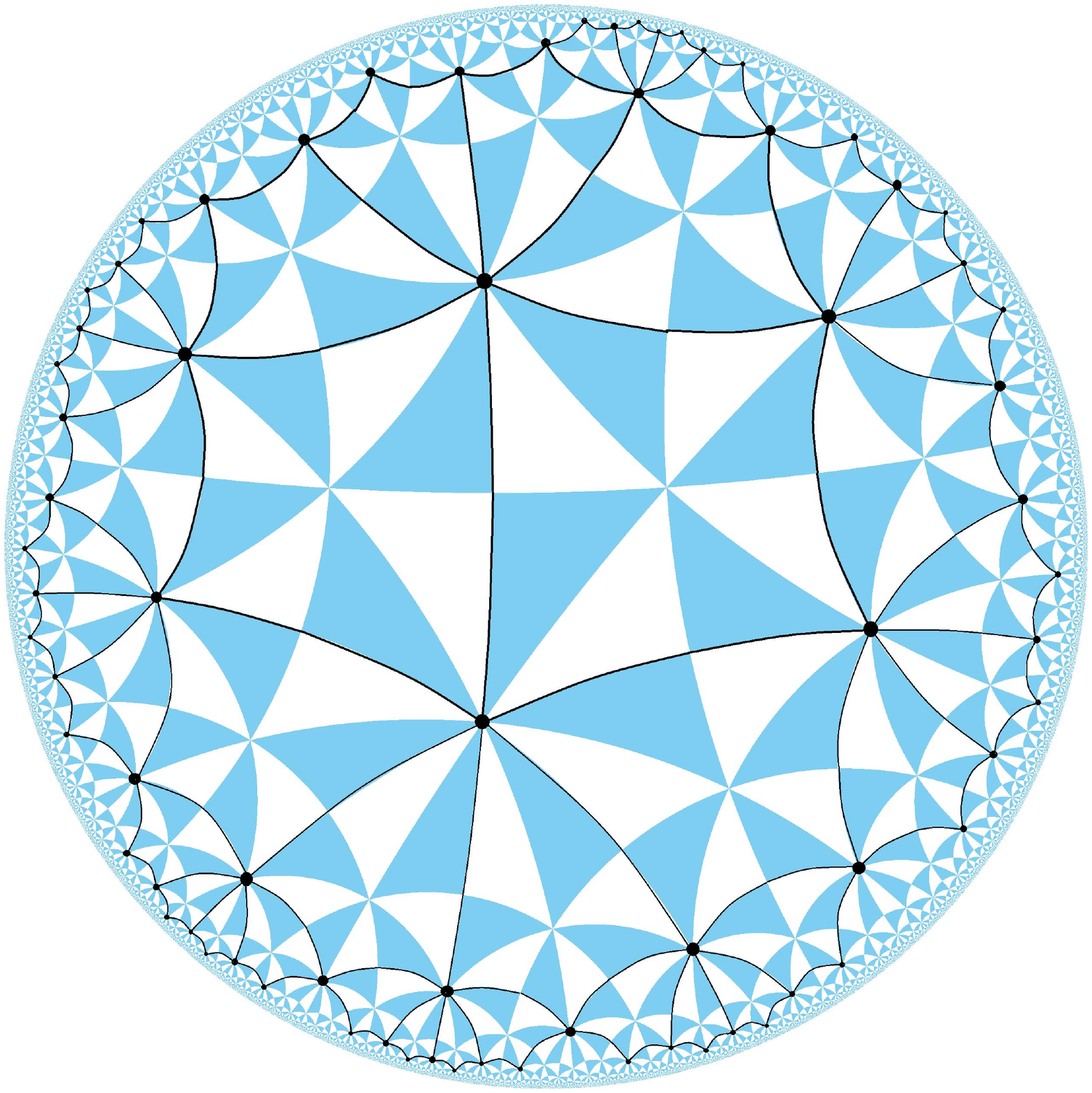}} 
\quad
\subfigure[][]{
\label{fig:H246tree}
\includegraphics[trim=0.5cm 5cm 0.5cm 4cm,width=0.3\textwidth]{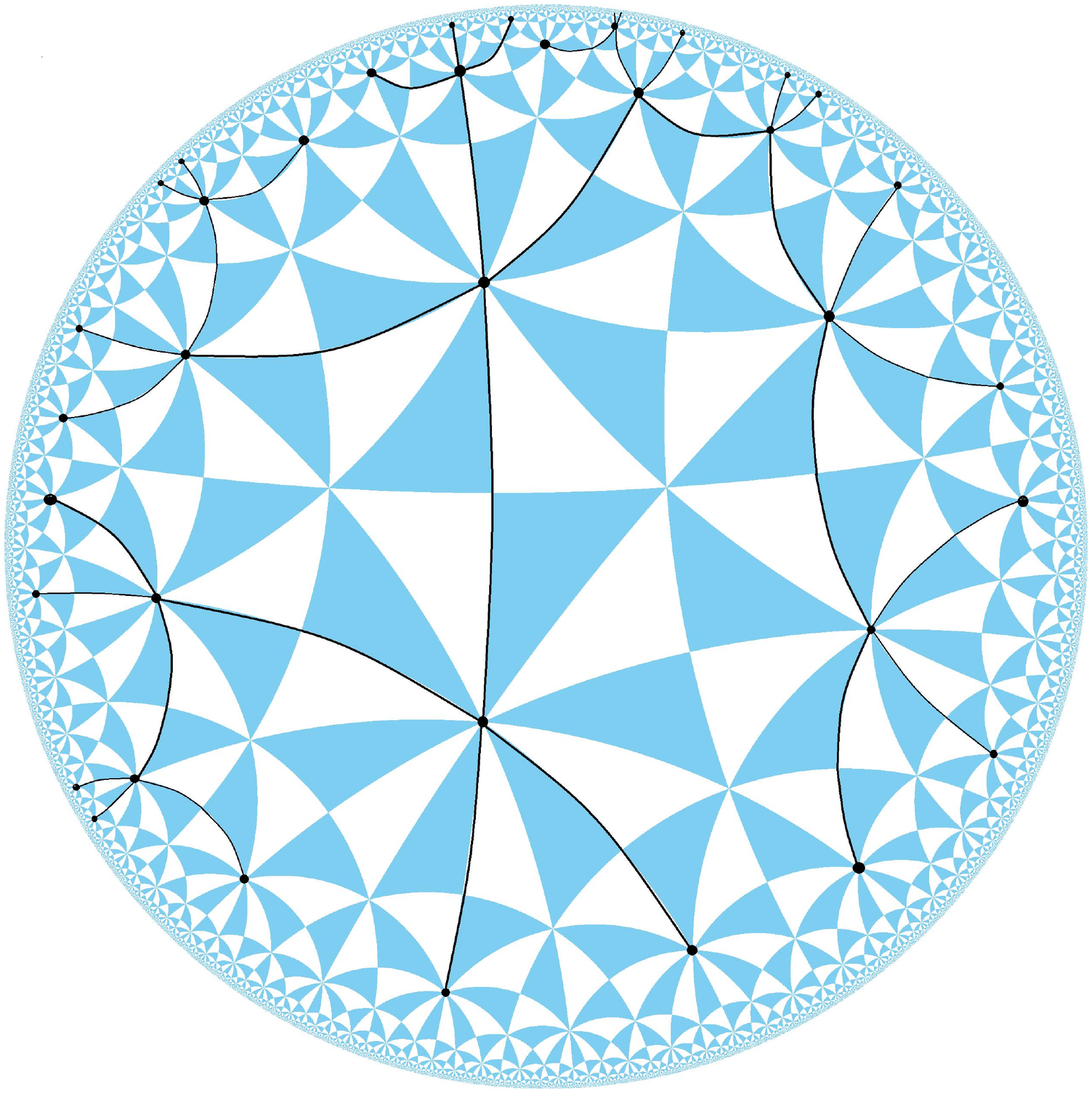}} 
\quad
\subfigure[][]{
\label{fig:tesswithtree}
\includegraphics[trim=0.5cm 5cm 0.5cm 4cm,width=0.3\textwidth]{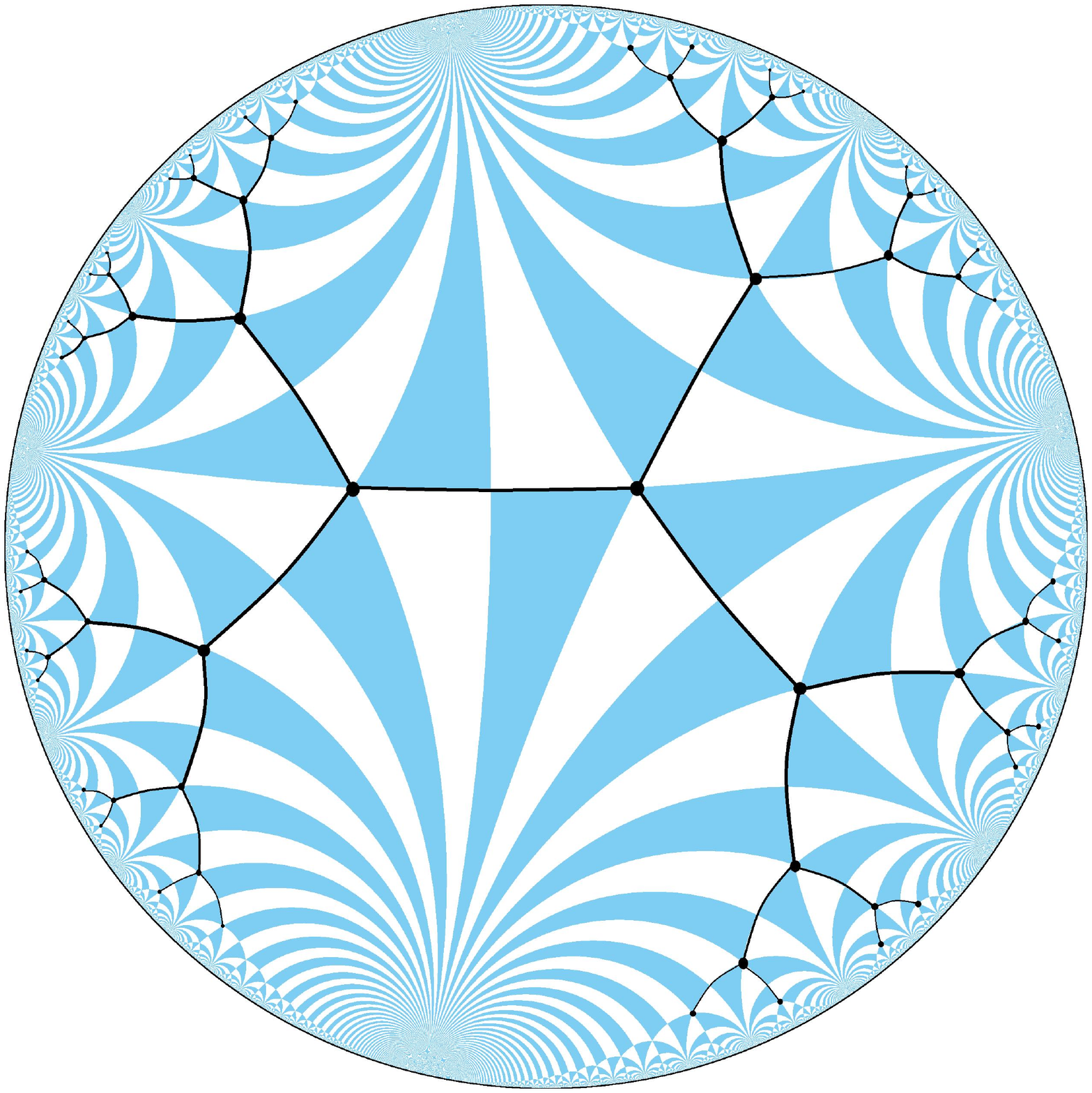}}
\caption{Tensor network embedded in a regular tessellation of Poincar\'e disk. 
(a): Hexagon tensor network, based on W$[2,4,6]$ tilling. 
(b): A $4$-valent tree tensor network, based on the spanning tree of W$[2,4,6]$ tilling. 
(c): $3$-valent tree tensor network, based on W$[\infty,2,3]$ tiling.}
\label{2tesse}
\end{figure}

For example, when the basis tile is a hyperbolic triangle with 
\begin{equation}
\textrm{Triangle }(\frac{\pi}{\ell}, \frac{\pi}{m},\frac{\pi}{n}) \qquad \textrm{with} \quad \frac{1}{\ell}+\frac{1}{m}+\frac{1}{n}<1
\end{equation}  
the isometry group of $\mathcal{G}$ is the triangle group $W[\ell,m,n]$. 
For generic types of tilings (made of basic triangles), the isometry group is a reflection group\footnote{Note that rotations can be generated by successive reflections across different edges. For textbooks on hyperbolic reflection groups see e.g.\ \cite{Humphreys}.} (or more abstractly Coxeter group), i.e.\ generated by reflections across the edges of the tiles. 
Figure \ref{fig:H246} shows the hexagon tensor network based on the W$[2,4,6]$ tilling. 
A brief review and relevant references can be found in \cite{Bhattacharyya:2016hbx}.
\smallskip

However, the representation theory of Coxeter groups is not strong enough to help find the bulk solutions (living on the graph $\mathcal{G}$). 
As a  contrast, in the usual AdS/CFT, the bulk solution can be solved explicitly in terms of irreducible representations of the conformal group. 
In particular, one can find the bulk solutions that are dual to primary operators of the boundary CFT. 
For tensor networks, to proceed to more quantitive comparisons between the bulk and the boundary, we need a better handle on the isometry group of the graph $\mathcal{G}$.\footnote{We are not aware of a systematic discussion of graph wavefunctions defined on regular tilings of the hyperbolic space, or if such solutions exist at all, how they organize themselves into representations of the Coxeter group. 
}

\subsubsection{Tensor networks on abstract tree}

The isometry of the tensor network based on a regular tiling is a \textit{discrete} subgroup of SL$(2,\mathbb{R})$ because we insist on the tensor network to be geometrically embedded in AdS$_2$, respecting its isometry. 
However, there is no a priori reason that the discretization should work in this most naive form.  
Suppose we relax this assumption, i.e.\ we view the lattice as an abstract lattice, free from the underlying AdS, then it is possible for the lattice to furnish a different, possibly bigger, symmetry. 
\smallskip

In this paper, we adopt the somewhat radical approach of giving up the geometric embedding in order to gain more symmetry. 
Since modeling AdS/CFT is our main goal, we still want this symmetry to be related to the conformal symmetry in some way.
It turns out that the $(p+1)$-valent tree with $p$ being a prime number can furnish a representation for the full conformal group of SL$(2,\mathbb{Q}_p)$ where $\mathbb{Q}_p$ is the field of $p$-adic numbers.
This is a continuous group and hence much bigger than a discrete subgroup of SL$(2,\mathbb{R})$. 
\smallskip

This consideration is inspired by the recently discussed $p$-adic AdS/CFT correspondence \cite{Gubser:2016guj, Heydeman:2016ldy}.
The proposed duality is a discrete analogue of AdS/CFT. 
In the simplest example where the bulk is two dimensional and the boundary one dimensional, the bulk geometry is given by the Bruhat-Tits (BT) tree, whereas the boundary is conjectured to be a theory that is defined on the field $\mathbb{Q}_p$ i.e.\ the $p$-adic numbers, and which preserves the SL$(2,\mathbb{Q}_p)$ symmetry.
\smallskip

Before moving on to a review of the $p$-adic tree, we emphasize that, different from \cite{Heydeman:2016ldy}, we do not view it as arising from a regular tessellation of the real AdS. 
There are two ways a tree can arise from tessellations. 
The first is the spanning tree of the graph of a tessellation --- we draw the spanning tree of the tessellation based on W$[2,4,6]$ in Figure \ref{fig:H246tree}.\footnote{Given a lattice, its spanning tree is a tree that contains all the vertices of the lattice and has the minimal number of edges.}
The second is when the triangle group is W$[\infty, 2, m]$, which results in a $m$-valent tree. 
Figure \ref{fig:tesswithtree} shows an example of the $3$-valent tree. 
\smallskip

However, the trees that arise from these two ways completely break the scaling symmetry of the underlying AdS space.\footnote{For a regular tessellation based on the triangle group W$[\ell,m,n]$ in which all three numbers are finite, there are some discrete scaling symmetries preserved by the tessellation. 
However, this is not enough for any quantitive calculation we want to do in this paper.} 
In contrast, the $p$-adic tree we will be considering furnishes the full SL$(2,\mathbb{Q}_p)$ symmetry.

\subsection{$p$-adic number field and Bruhat-Tits tree}

In this section we review the $p$-number field $\mathbb{Q}_p$ and its Bruhat-Tits tree, to prepare for the discussion of the Bruhat-Tits tree and $p$-adic AdS/CFT, and to fix notation. 
For textbooks on $p$-adic numbers, see \cite{Koblitz, Gouvea}. 
For its applications in string theory or other fields of mathematical physics, we recommend \cite{Freund:1987kt,Brekke:1993gf,Brekke:1988dg, Freund:1987ck,Manin:2002hn,Dragovich:2007wb,Dragovich:2009hd}.

\subsubsection{The field $\mathbb{Q}_p$ of $p$-adic numbers}

The field of rational numbers $\mathbb{Q}$ can be extended to the field  of real numbers $\mathbb{R}$, with respect to the Euclidean norm $|x|$, which satisfies a few axioms. 
\begin{equation}
\begin{aligned}
&(1):\,\, |x|\geq 0 \qquad (2):\,\, |x|=0 \leftrightarrow x=0\qquad (3):\,\,  |x\, y|=|x|\, |y| \quad \forall x,y \in \mathbb{R}\\
&(4):\,\, |x+y| \leq |x|+ |y|  \qquad \forall  x,y \in \mathbb{R} \quad \textrm{(triangle inequality)}
\end{aligned}
\end{equation}

Starting with the rational field $\mathbb{Q}$, it is possible to extend it in other ways, with respect to different norms that obey the above axioms.  
\smallskip

Given a prime number $p$, a rational number $x\in\mathbb{Q}$ can be uniquely expanded in terms of powers of $p$:\begin{equation}\label{xpexpansionQ}
x_p=\sum^{\infty}_{n=-N} a_n \, p^n \qquad\textrm{with} \qquad a_n \in \mathbb{F}_p 
\end{equation}
where $\mathbb{F}_p$ denotes the residue field consisting of integers $0,1,\dots, p-1$.
The expansion (\ref{xpexpansionQ}) can be rewritten to 
highlight  the congruence of $x_p$ with respect to $p$, i.e.\
the leading term of the $p$-adic expansion:
\begin{equation}
x= p^{v_p(x)} \sum^{\infty}_{n=0} b_n \, p^n 
\qquad  \textrm{with} \quad b_0 \neq 0\,, \quad b_n \in \mathbb{F}_p \quad \textrm{and}\quad v_{p}(x) \in \mathbb{Z}
\end{equation}
using which the $p$-adic norm $|x|_p$ of $x$ is defined as 
\begin{equation}\label{padicnorm}
|x|_p \equiv p^{-v_p(x)} \qquad \textrm{with}\quad v_{p}(x) \in \mathbb{Z}.
\end{equation}
Namely, the more divisible $x$ is w.r.t. $p$, the smaller norm it has. \smallskip 

It is then easy to check that the $p$-adic norm obeys all four axioms for the norm.
In fact, it satisfies an even stronger form of the fourth axiom:\footnote{Note that the original triangle inequality is trivially satisfied by the p-adic norm:
$|x+y|_p \leq |x|_p +|y|_p$.
} 
\begin{equation}\label{STI}
|x+y|_p \leq \textrm{max}(|x|_p,|y|_p) \qquad \textrm{(strong triangle inequality)}
\end{equation}

We thus see that the  rational field $\mathbb{Q}$ can have infinitely many different norms: the Euclidean norm $|x|$ together with the  $p$-adic norms $|x|_p$ for each prime $p$.\footnote{The Euclidean norm $|x|$ and the $p$-adic norm $|x|_p$ are the only possible norms to complete the rational field $\mathbb{Q}$ (giving $\mathbb{R}$ and $\mathbb{Q}_p$, respectively), as already shown by  Ostrowski in 1919 \cite{Ostrowski}.} 
The real field $\mathbb{R}$ is only one possible extension of $\mathbb{Q}$, using the Euclidean norm $|x|$. 
Now, for each prime $p$, we can have a different extension of $\mathbb{Q}$ using the  $p$-adic norms $|x|_p$.
Given a fixed prime number $p$, the field $\mathbb{Q}_p$ consists of all possible formal expansions of the form:
\begin{equation}\label{QpDef}
\mathbb{Q}_p \equiv \{x_p=\sum^{\infty}_{n=-N} a_n \, p^n \,\, |\,\,a_n \in \mathbb{F}_p\}.
\end{equation}
The $p$-adic norm (\ref{padicnorm}), in particular $|p^n|_p=\frac{1}{p^n}$, ensures that  the formal series (\ref{QpDef}) converges.\footnote{Note that  in contrast to the decimal expansion for the real number $x\in\mathbb{R}$, for a $p$-adic number $x_p\in \mathbb{Q}_p$, we allow the expansion to be infinite in the direction of the positive exponent of $p$ but not along the negative direction, precisely because a higher power of $p$ has a smaller $p$-adic norm.} The strong triangle inequality (\ref{STI}) also implies $|x+x|_p \leq |x|_p$, which violates the Archimedes principle $|x+x| \geq |x|$ --- hence the geometry based on $p$-adic norm is called non-Archimedean.
\smallskip

The $p$-adic norm is  used to define the following subset of $\mathbb{Q}_p$, which will be useful in the later construction of the Bruhat-Tits tree and the discussion on $p$-adic integration.  
First, the \textit{unit sphere in $\mathbb{Q}_p$} consists of $x_p$ with unit norm:
\begin{equation}\label{unitsphere}
\mathbb{U}_p\equiv \{x_p \in\mathbb{Q}_p \, |\, |x|_p=1 \} 
\qquad i.e. \quad x|_p= a_0+ a_1 p +a_2 p^2 +\dots \qquad a_0\neq 0
\end{equation}
The \textit{unit ball of $\mathbb{Q}_p$} is inside $\mathbb{U}_p$:
\begin{equation}\label{unitball}
\mathbb{Z}_p\equiv \{x_p \in\mathbb{Q}_p |\, |x|_p\leq 1 \}     
\qquad i.e. \quad       x|_p=a_0+ a_1 p +a_2 p^2 +a_3 p^3\dots
\end{equation}
Note that the unit ball $\mathbb{Z}_p$ is precisely the ring of $p$-adic integers. 
However, unlike $\mathbb{Z}$ (which is open in $\mathbb{R}$), $\mathbb{Z}_p$ is both open and closed (``clopen") in $\mathbb{Q}_p$. 
Finally, we denote the set of non-zero elements in $\mathbb{Q}_p$ as
$\mathbb{Q}^{*}_p\equiv \mathbb{Q}_p/\{0\}$, which is
$\mathbb{Q}^{*}_p \equiv \coprod_{n\in\mathbb{Z}} p^n \, \mathbb{U}_{p}$.

\subsubsection{Bruhat-Tits tree as bulk of $p$-adic line $\mathbb{Q}_p$}
In this subsection we summarize the construction of the Bruhat-Tits tree \cite{BruhatTits,Zabrodin:1988ep},
in particular motivating it from its role as the bulk of $\mathbb{Q}_p$ (i.e.\ the analogue of upper half plane but whose boundary is $\mathbb{Q}_p$ instead of $\mathbb{R}$) and prepare for the discussion on the SL$(2,\mathbb{Q}_p)$ action on the tree. 
\smallskip

The real field $\mathbb{R}$ is the boundary of the upper half plane $\mathbb{H}\equiv \textrm{SL}(2,\mathbb{R})$/SO$(2,\mathbb{R})$. 
With coordinates $\mathbb{H}\equiv\{z=x+i\,y\, |\, x\in \mathbb{R}, y\in\mathbb{R}_+\}$, it has SL$(2,\mathbb{R})$-invariant metric $ds^2=\frac{1}{y^2}(dx^2+dy^2)$. 
An SL$(2,\mathbb{R})$ action on a point $z=x+i\, y$ on $\mathbb{H}$ would induce the same action on its boundary point $x$. 
If we replace the boundary space $\mathbb{R}$ by the $p$-adic field $\mathbb{Q}_p$, what would be its bulk, i.e. what is the $p$-adic version of the upper half plane?
\smallskip

The analogy with the relation between $\mathbb{H}$ and its boundary $\mathbb{R}$, shown in Table \ref{tableBBrelations} 
\begin{table}
\begin{center}
        \begin{tabular}{|c|c|c|}
                \hline
                &       upper half plane $\mathbb{H}$ & Bruhat-Tits tree $\mathbb{H}_p$\\
                \hline
                Isometry group $G$      & SL$(2,\mathbb{R})$  & PGL$(2,\mathbb{Q}_p)$    \\
                Isotopy group $K$ & SO$(2,\mathbb{R})$& PGL$(2, \mathbb{Z}_p)$\\
                Boundary & $\mathbb{R}$ &  $\mathbb{Q}_p$ \\    
                \hline
        \end{tabular}
        \caption{The parallel between the upper half plane $\mathbb{H}$ and the Bruhat-Tits tree $\mathbb{H}_p$.}
        \label{tableBBrelations}
        \end{center}
\end{table}
suggests that one should replace $\mathbb{R}$ by $\mathbb{Q}_p$ in the definition of $\mathbb{H}$ to give
\begin{equation}\label{Hcoset}
\mathbb{H}_p \equiv \frac{\textrm{PGL}(2,\mathbb{Q}_p)}{\textrm{PGL}(2,\mathbb{Z}_p)}.
\end{equation}
We immediately see the difference from the real case.
As the maximal compact subgroup of the isometry group $\textrm{PGL}(2,\mathbb{Q}_p)$,  $\textrm{PGL}(2,\mathbb{Z}_p)$ is both open and closed (``clopen") in  $\textrm{PGL}(2,\mathbb{Q}_p)$. 
Therefore, although $\mathbb{Q}_p$ is a continuum, its bulk $\mathbb{H}_p$ is actually \textit{discrete}.

\smallskip

Since $\mathbb{H}_p$ has a discrete topology, we cannot simply give $\mathbb{H}_p$ the coordinate $z_p=x_p + i\, y_p$ in which $x_p \in \mathbb{Q}_p$ and $y_p \in \mathbb{Q}_{p+}$ and write down the PGL$(2,\mathbb{Q}_p)$ invariant metric on it. 
However, the coset expression (\ref{Hcoset}) suggests that one can construct it  as a set of equivalence classes  $\langle \langle \vec{f},\vec{g} \rangle \rangle$ of lattices $\langle \vec{f},\vec{g} \rangle $ in $\mathbb{Q}_p\otimes \mathbb{Q}_p$, where two lattices are equivalent, $\langle \vec{f},\vec{g} \rangle \sim \langle \vec{f}',\vec{g}' \rangle $, iff 
\begin{equation}\label{isotropy}
        \begin{aligned}
                ( \vec{f}',\vec{g}' ) 
                =  ( \Gamma \cdot \vec{f}, \Gamma \cdot \vec{g}  ) 
                 \qquad \textrm{with}\quad \Gamma
                  \in \textrm{PGL}(2, \mathbb{Z}_p ).
        \end{aligned}
\end{equation}
We leave the details of the construction to the appendix. To summarize, the $p$-adic analogue of the upper half plane $\mathbb{H}_p$ has the topology of  an infinite $(p+1)$-valence tree (called Bruhat-Tits tree).
The nodes on the tree are defined as equivalence classes  $\langle \langle \vec{f},\vec{g} \rangle \rangle$ of lattices $\langle \vec{f},\vec{g} \rangle $ and have the form
\begin{equation}\label{Hpcoord}
        \langle \langle \left(\begin{matrix}p^m\\ 0 \end{matrix}\right), \left(\begin{matrix}x^{(m)}\\ 1\end{matrix}\right)\rangle\rangle
  \qquad \qquad x^{(m)}=\sum^{m-1}_{ n=-N} a_n p^n \qquad a_n \in\mathbb{F}_p.   \end{equation}
Note that since $x^{(m)}$ truncates at $p^{m}$, we can think of $p^m$ as giving the accuracy level of a $p$-adic number $x^{(m)}$, i.e.\ the node (\ref{Hpcoord}) represents the \textit{equivalence class} 
\begin{equation}
x^{(m)}+p^m \mathbb{Z}_p.
\end{equation}

This somewhat formal definition of the Bruhat-Tits tree as equivalence classes  of lattices  in $\mathbb{Q}_p\otimes \mathbb{Q}_p$ actually connects nicely with the $p$-adic expansion of the boundary $\mathbb{Q}_p$.
First note that the $p$-adic expansion already has a natural tree structure. Consider a generic $p$-adic number
\begin{equation}\label{xpexpansion}
\dots \, a_3 \, a_2\, a_1 \, a_0\, . \, a_{-1}\,a_{-2} \, a_{-3}\, \dots a_{-N}
\end{equation}
First, start from level $p^0$, there are $p$ choices for the coefficient $a_{0}$, draw a node for each choice. The node with $a_0=0$ is then the origin $O$.
Starting from each node (labeled by $a_{0}$) at level $p^{0}$, there are again $p$ choices for $a_{1}$ --- draw these $p$ nodes at level $p^{1}$ and connect them to the node they start from.
Moving up this way (and also connecting all nodes at $p^0$ backwards to the node the correspond to $0 p^{-1}+0$), one draws an infinite $(p+1)$-valent tree starting from level $p^{-1}$. 
Moving backwards towards negative powers $p^{-n}$ then  gives the entire Bruhat-Tits tree.  
\smallskip

Thus we obtain a one-to-one map between a $p$-adic number and a branch on the Bruhat-Tits tree: given a $p$-adic number, its branch is defined by starting from the lowest power of the  expansion (\ref{xpexpansion}) and then at each level $p^n$ following the twig corresponding to $a_n$ in the expansion (\ref{xpexpansion}).
Each node in the bulk Bruhat-Tits tree has two label: \begin{equation}\label{BTnode}
z=x^{(m)} \qquad \qquad  z_0=p^m
\end{equation}
where $p^m$ gives the accuracy level and $x^{(m)}$
a $p$-adic number to the accuracy $p^m$, i.e. it represents the equivalence class $x^{(m)}+p^m \mathbb{Z}_p$. 
This precisely agrees with the result from the lattice construction of the Bruhat-Tits tree.
\smallskip

The non-zero elements in $\mathbb{Q}_p$ can be grouped according to the leading term in the $p$-adic expansion:
\begin{equation}
\mathbb{Q}^{*}_p = \coprod_{n\in\mathbb{Z}} p^n \, \mathbb{U}_{p}: \qquad 
x|_p=p^{v} \, u \qquad \textrm{with}\qquad v\in \mathbb{Z}\qquad  \textrm{and}\qquad u\in \mathbb{U}_p
\end{equation}
i.e. $u=\sum^{\infty}_{n=0}a_n \, p^n$ with $a_0\neq 0$. 
The set $p^n \mathbb{U}_p$ for each $n\in \mathbb{Z}$ forms a subtree with the root at:
\begin{equation}
\textrm{Points on the main branch:} \qquad x^{(n)}_0=p^n \cdot 0 \qquad \textrm{with accuracy}\qquad p^n
\end{equation}
The line connecting all the $x^{(n)}_0$ is then the main branch, running from $n\rightarrow - \infty$ to $n\rightarrow +\infty$.

\subsection{Conformal primaries for tensor network on Bruhat-Tits tree}

Recall that the main disadvantage of viewing the network $\mathcal{G}$ as a naive discretization of the substrate AdS space is that it retains too few symmetries and makes it difficult to organize operators. 
We now show that identifying the tree network $\mathcal{G}$ as the Bruhat-Tits tree preserves a full conformal SL$(2,\mathbb{Q}_p)$ symmetry\footnote{Here we are a bit cavalier in our notation: the actual group that acts is $\textrm{PGL}(2,\mathbb{Q}_p)$, but we shall often (in analogy with the table above) refer to it as SL$(2,\mathbb{Q}_p)$.}  for the tensor network, and in particular allows us to define conformal primaries for operators acting on the tensor network. 

\subsubsection{SL$(2,\mathbb{Q}_p)$ action on Bruhat-Tits tree}
 \label{sec:scaling}
 
The coordinate system (\ref{Hpcoord}) assigns a vertex on the Bruhat-Tits tree two numbers $p^m$ and $x^{(m)}$. 
In order to determine the radial direction and the boundary direction, and how to choose the cutoff surface in a manner suitable to holography (and analogous to the upper half plane in the real case), we now study their behavior under the bulk PGL$(2,\mathbb{Q}_p)$ action.
\smallskip

A $ \textrm{PGL}(2, \mathbb{Q}_p )$ transformation acts on the lattice via
\begin{equation}
\langle \vec{f},\vec{g}\rangle  \rightarrow 
\langle\gamma \cdot \vec{f}, \gamma \cdot \vec{g} \rangle \ \qquad \textrm{with}\qquad \gamma=\left(\begin{matrix}a&b\\ c &d\end{matrix}\right) \in \textrm{PGL}(2,\mathbb{Q}_p).
\end{equation} 
Given a vertex on the Bruhat-Tits tree with coordinate (\ref{Hpcoord}), under a PGL$(2,\mathbb{Q}_p)$ action, it transforms as
\begin{equation}
\begin{aligned}
\langle \langle 
\left(\begin{matrix}p^m\\ 0 \end{matrix}\right), \left(\begin{matrix}x^{(m)}\\ 1\end{matrix}\right)\rangle\rangle 
\qquad \longrightarrow \qquad 
\langle \langle \label{eq:treetrans}
\left(\begin{matrix} p^{m'}\\ 0 \end{matrix}\right), \left(\begin{matrix}\frac{a\,x^{(m)}+b}{c\,x^{(m)}+d}\\ 1\end{matrix}\right) 
\rangle\rangle 
\end{aligned}   
 \end{equation}
where
\begin{equation}\label{padicSL2y}
p^{m'}= p^{m} \left|\frac{(c\,x+d)^2}{a d- b c} \right|_p  . 
\end{equation}
Namely, start with the bulk point $x=\sum^{m-1}_{ n=-N} a_n \,p^n$ 
, with accuracy only up to level $p^m$, its SL$(2,\mathbb{Q}_p)$ image is another bulk point at
\begin{equation}
\frac{a\,x^{(m)}+b}{c\,x^{(m)}+d} =\sum^{m'-1}_{ n=-N} b_n \,p^n \qquad \quad \textrm{with accuracy } \quad p^{m'} .
\end{equation}

\begin{figure}[h!]
        \centering
        \includegraphics[trim=0.5cm 9cm 0.5cm 4cm, width=0.7\textwidth]{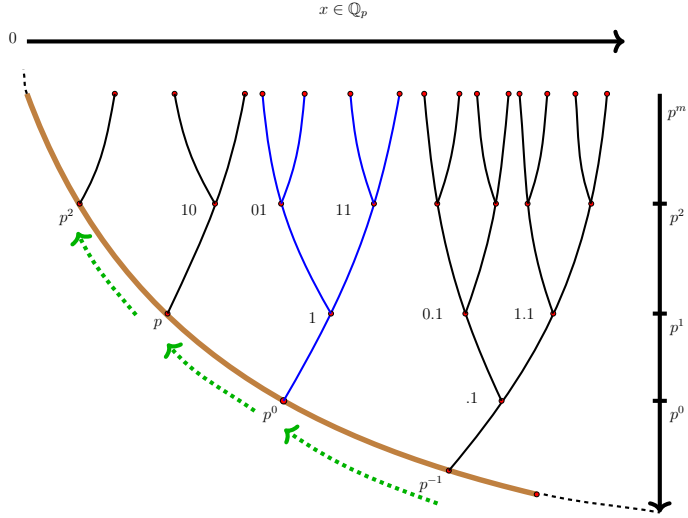}
        \caption{Bruhat-Tits tree for $p=2$. The dilatation $D$ on the tree slides the branches along   the main branch.         }
        \label{tess2}
\end{figure}

It is quite remarkable that the Bruhat-Tits tree, though discrete, can furnish the full conformal group PGL$(2,\mathbb{Q}_p)$. 
This allows us to study the function on the tree which has  a definite quantum number. 
Given the Iwasawa decomposition $G=NAK$, where $N$ is the Borel subgroup, $A$ the dilation, and $K$ maximal compact subgroup SL$(2,\mathbb{Z}_p)$, it is enough to look at their actions separately. 
The most important is the scaling transformation. 
\smallskip
Under a dilatation by $p^{n}$, a vertex on the tree transforms as 
        \begin{equation}\label{eq:dilatation}
D=      \left(\begin{matrix}p^{n/2} &0
        \\0&p^{-n/2} \end{matrix}\right):
\qquad  \langle \langle \left(\begin{matrix}p^m\\ 0 \end{matrix}\right), \left(\begin{matrix}x^{(m)}\\ 1\end{matrix}\right)\rangle\rangle \,\, \longrightarrow \,\,
\langle \langle \left(\begin{matrix} p^{m+n}\\ 0 \end{matrix}\right), \left(\begin{matrix} p^{n}x^{(m)}\\ 1\end{matrix}\right) \rangle\rangle .
  \end{equation}
The action moves the branches along the main branch, shown in Figure \ref{tess2}.

\subsubsection{Choice of cutoff surface}

The construction of a holographic correspondence includes a prescription on how the boundary is approached from inside the bulk, i.e.\ how to define the cut-off surface which is then pushed to infinity. 
For instance, AdS in global coordinates or Poincar\'e coordinates have different natural cut-off surfaces and therefore different boundary behaviors. 
Now we show that the cut-off surface natural to the Bruhat-Tits tree should be lines of constant $p^m$.
\smallskip

As we go to the boundary, both $m$ and $m'$ $\rightarrow \infty$, 
\begin{equation}\label{eq:xmcoordinate}
x^{(m)} \, =\sum^{m-1}_{ n=-N} a_n \,p^n\quad
\longrightarrow \quad x\, =\sum^{\infty}_{ n=-N} a_n \,p^n
\end{equation}
we have the expected boundary $ \textrm{PGL}(2, \mathbb{Q}_p )$ transformation:
\begin{equation}\label{padicSL2x}
x\, =\sum^{\infty}_{ n=-N} a_n \,p^n \ \qquad \longrightarrow \qquad  \frac{a\,x+b}{c\,x+d} \, =\sum^{\infty}_{ n=-N'} b_n \,p^n
\end{equation}

\begin{figure}[h!]
        \centering
        \includegraphics[ width=0.5\textwidth]{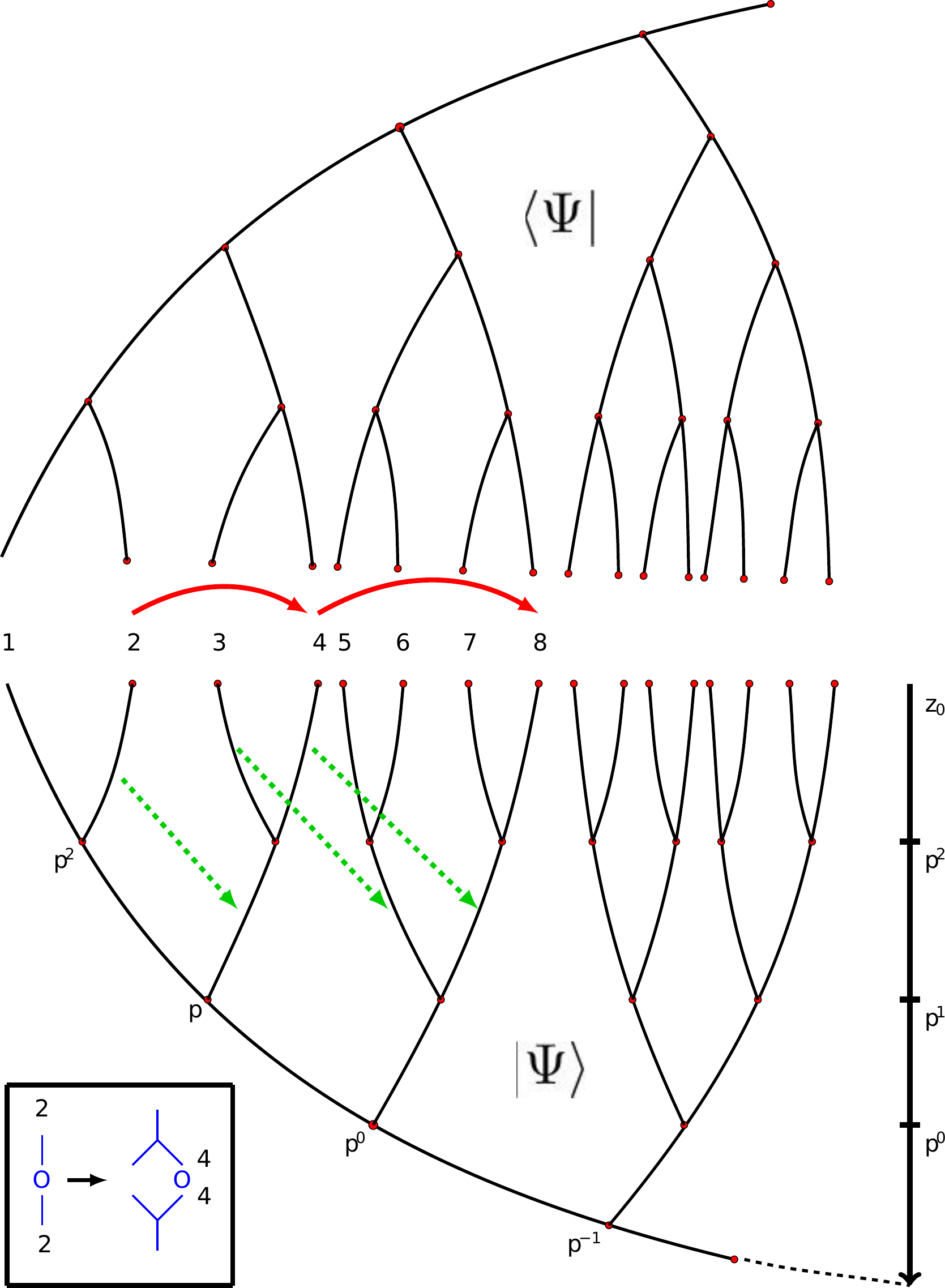}
        \caption{A tree tensor network together with its conjugate, glued together (in computation of correlators)  along the common boundary, which is the analogue of the Poincar\'e cutoff surface in the BT tree. 
Tensors move down the tree under a scaling transformation. 
We indicate with green arrows a few examples of how the links move. 
Such a transformation corresponds to a transformation of operators such that it is sandwiched between a network tensor. This is depicted in the box. 
}
        \label{fig:treecutoff}
\end{figure}

\smallskip
Let's compare to the real case. 
Under an SL$(2,\mathbb{R})$  on a point $z= x+ i\, y$ in  the upper half plane $\mathbb{H}$,  $x$ and $y$ transform as 
\begin{equation}\label{eq:coordtrans}
\begin{aligned}
x &\quad\longrightarrow \quad\frac{(a\,x+b)(c\,x+d) +a\,c \,y^2}{(c \,x +d)^2 +(c\, y)^2}\quad \xlongrightarrow{y\rightarrow 0} \quad \frac{a\,x+b}{c\, x +d } \\
y &\quad \longrightarrow \quad\frac{y}{(c\, x +d)^2 +(c \,y)^2}\qquad \qquad\,\xlongrightarrow{y\rightarrow 0} \quad \frac{y}{(c\, x +d)^2 } 
\end{aligned}
\end{equation}
Comparing this with the transformation of $x^{(m)}$ in (\ref{padicSL2x}) and $p^{m}$ in (\ref{padicSL2y}), we conclude that $p^{m}$ should play the role of the holographic direction, the analogue of $y$ in the real case, whereas $x^{(m)}$ the role of the boundary direction.
This also means that the cut-off surface should be a line of constant $p^m$, as shown in Figure \ref{fig:treecutoff}. 
This is analogous to the choice of the $z=\epsilon$ surface as the cut-off surface, where $z$ is the radial direction in Poincar\'e coordinates.

\subsubsection{Conformal primaries for $p$-adic tensor network}
\label{sec:conformalprimary1}

A primary field $\mathcal{O}^{I}(x)$ of PSL$(2,\mathbb{Q}_p)$ with conformal weight $\Delta_I$ is defined as \cite{Melzer:1988he} 
\begin{equation}\label{primary}
O^I(\frac{a\, x +b}{ c\, x+d} ) =\left( \left| \frac{\det\gamma}{(c \, x +d)^2  }\right|_p \right)^{-\Delta_I} O^{I}(x) 
\qquad \textrm{where}\quad
\gamma=\left(\begin{matrix}a&b\\ c &d\end{matrix}\right) \in \textrm{PGL}(2,\mathbb{Q}_p)
\end{equation}
In particular, for a scaling transformation $x\rightarrow p\,x$
\begin{equation}\label{primaryp}
O^I(px ) =p^{\Delta_I} O^{I}(x) 
\end{equation}
Let's translate this condition on operators in the tensor network.
\smallskip

Consider an operator $\mathcal{O}^I$ acting on a boundary leg of the tensor network.
One important feature of the constant $p^m$ cutoff surface is that the external legs are not evenly distributed, see Figure \ref{fig:treecutoff}.
The distance of the external legs from the main-branch increases as we move along the cutoff surface. 
Therefore, under the scaling transformation $x\rightarrow p^n \, x$ with $n>0$, an operator acting on a given boundary leg on the cutoff surface would hop from a branch closer to the main branch to branches further away from the main branch. 

For instance, under the boundary scaling $x\rightarrow p\,x$, an operator $\mathcal{O}^I$ acting on leg 2 is mapped to $\mathcal{O}^I$ acting on leg 4, and leg 4 to leg 8, and so on. Applying (\ref{primary}) to this particular network, we have
\begin{equation}\label{hop}
\mathcal{O}^I(i_{2})=p^{-\Delta_I} \mathcal{O}^I(i_{4}) 
\end{equation}

As we will see later in Section 6, the condition such as (\ref{hop}), together with the assumption on the homogeneity of the network, allows us to explicitly construct a primary basis starting from the Pauli basis defined in (\ref{eq:genPaul}), and moreover relate it to the ``operator pushing" basis defined in (\ref{HKLLemergeD}).

\section{$p$-adic HKLL from tree tensor networks}
 \label{sec:padicbulkrec}

In section \ref{sec:bulkrec} we derived the bulk operator reconstruction formula for generic tensors using ``operator pushing". 
In this section we show that for tree tensor networks, it matches nicely with the expectation from the conjectured $p$-adic AdS/CFT correspondence.

\subsection{HKLL for tree tensor network}

\begin{figure}[!h]
        \centering
        \includegraphics[trim=0cm 13.5cm 0cm 3.5cm, width=0.7\textwidth]{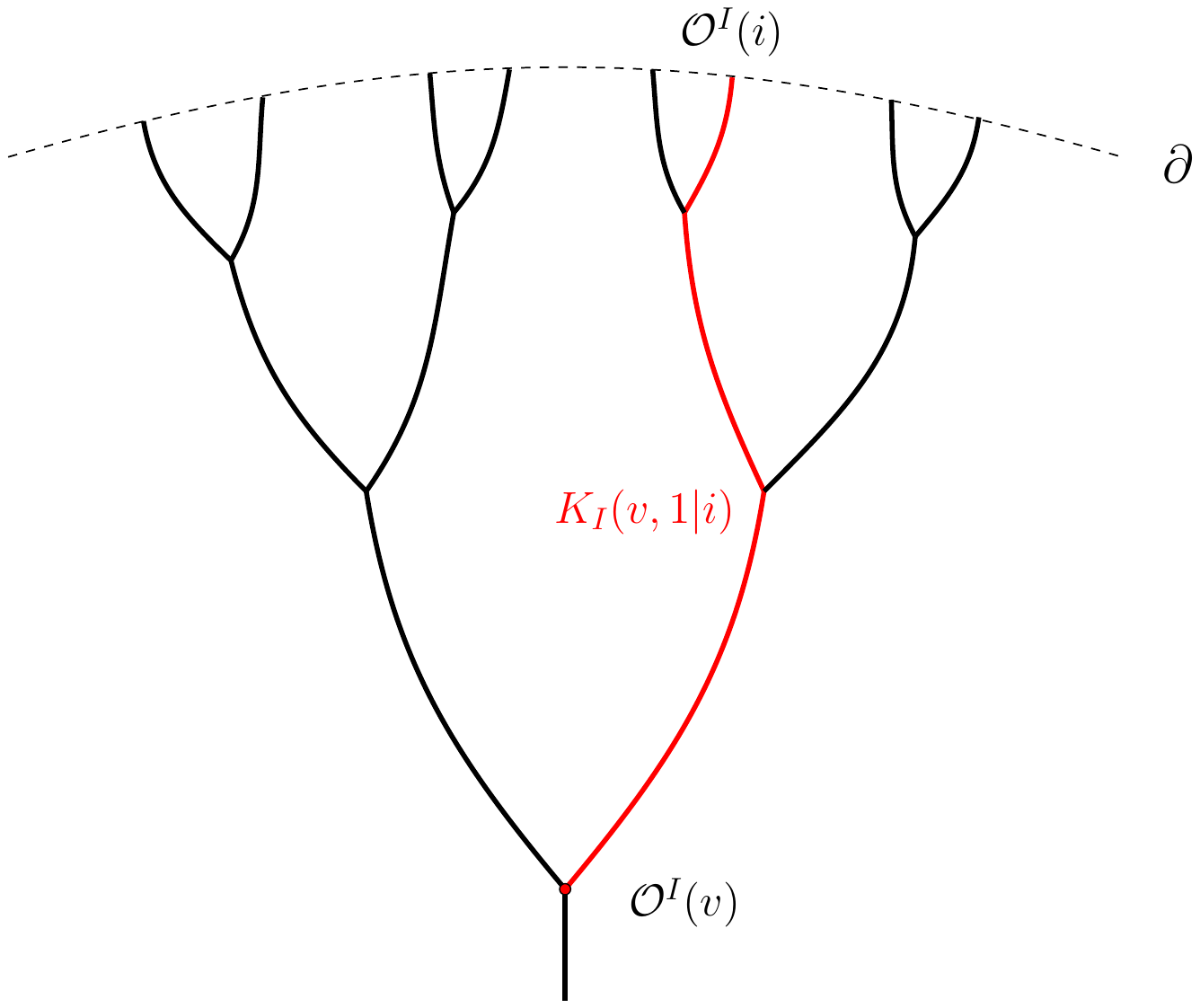}
        \caption{Linear contribution of global operator pushing for a tree tensor network.}
        \label{fig:linearHKLL}
\end{figure}

When the network $\mathcal{G}$ is an $r=p+1$ valent tree, the path from a bulk edge (say the first edge of vertex $v$) to the boundary edge is unique. 
Therefore the linear term of the global operator pushing (\ref{HKLLemergeF}) becomes
\begin{equation}\label{LinearGlobalPushing}
\mathcal{O}^I(v,1) \, |\Psi_{\textrm{bulk}}\rangle =\sum_{i} (\lambda_I)^{|\mathcal{P}(v\rightarrow i)|} \mathcal{O}^I(i)\, |\psi_{\textrm{bndy}}\rangle
\end{equation}
where $\mathcal{P}(v\rightarrow i)$ labels the unique path from the bulk vertex $v$ to the boundary edge $i$, and $|\mathcal{P}|$ measures its length. 
This is illustrated in Figure \ref{fig:linearHKLL}.
\smallskip

It is exactly in the form of the HKLL formula  
\begin{equation}
\mathcal{O}^I(v,1) \, |\Psi_{\textrm{bulk}}\rangle =\sum_{i} K_A (v,1|i)\mathcal{O}^I(i)\, |\psi_{\textrm{bndy}}\rangle
\end{equation}
with the ``smearing function"
\begin{equation}\label{smearTTN}
K_{I}(v,1|i) \equiv (\lambda_I)^{|\mathcal{P}(v\rightarrow i)|} = p^{-\sigma_I |\mathcal{P}(v\rightarrow i)|} 
\end{equation}
where in the last step we have used the following definition
\begin{equation}\label{sigmadef}
p\equiv r-1 \qquad \textrm{and} \qquad  \sigma_I\equiv  - \frac{\ln \lambda_{I}}{\ln p} 
\end{equation}
\smallskip

\begin{figure}[!h]
        \centering
       \includegraphics[trim=1cm 13.5cm 1cm 4cm, width=0.7\textwidth]{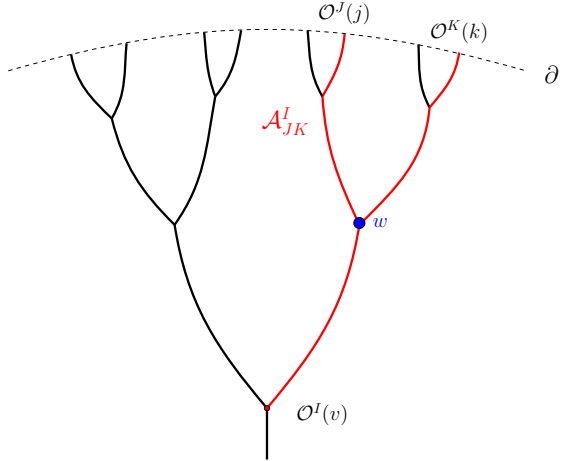}
        \caption{Non-linear contribution of global operator pushing for a tree tensor network.
         }
        \label{fig:nonlinearHKLL}
\end{figure}

Similarly, for the next order in the HKLL formula (\ref{nonlinearA}), 
the bulk-bulk kernel defined in (\ref{TNbulkG}) is also greatly simplified for a tree tensor network:
\begin{equation}\label{TTNbulkG}
G_{I}(v|w) \equiv (\lambda_I)^{|\mathcal{P}(v\rightarrow w)|} = p^{-\sigma_I |\mathcal{P}(v\rightarrow w)|} 
\end{equation}
Note that (\ref{nonlinearA}) is a sum of bifurcated paths from vertex $v$ to the two boundary legs $j$ and $k$, with the sum over the bifurcating point $w$, as illustrated in Figure \ref{fig:nonlinearHKLL}.

In both the ``smearing function" (\ref{smearTTN}) and the bulk-bulk kernel (\ref{TTNbulkG}) for an operator $\mathcal{O}^I$, only one parameter $\sigma_I$ (\ref{sigmadef}) appears. What is its physical meaning?
\smallskip

To answer this, one first checks that both the ``smearing function" (\ref{smearTTN}) and the bulk-bulk kernel (\ref{TTNbulkG}) satisfy the expected EOM for a tree Lagrangian of a scalar \cite{chungbook}: 
\begin{equation} \label{Greenfn}
(\Box_{v} + m_I^2)\, G_I(v|w) =   \mathcal{N}_I \delta(v,w) \qquad \qquad (\Box_{v} + m_I^2)\, K_I(v|i) =  0
\end{equation}
where the graph Laplacian $\Box$ can be defined for a generic graph as\footnote{Note the opposite sign from the usual definition --- this is to match the mostly-negative signature.}
\begin{equation} \label{eq:graphlap}
\Box\, \phi(v) = \sum_{v'\in \textrm{nn}(v)} \left(-\phi(v') +  \phi(v)\right)
\end{equation}
The sum is over all nearest neighbours of $v$.
We see that the parameter $\sigma_I$ is related to the mass squared of the bulk scalar by 
\begin{equation}\label{m2HKLL}
m_I^2=p^{1-\sigma_I}+p^{\sigma_I}-(p+1) .
\end{equation}

For a finite network $\mathcal{G}$, the ``smearing function" can be simply obtained by taking the second bulk vertex in the bulk-bulk kernel all the way to the boundary:
\begin{equation}
K_{I}(v,1|i) =\textrm{lim}_{w \rightarrow i} G_I(v|w).
\end{equation} 
For an infinite network, a regularization is needed to make $K_{I}(v,1|i)$ finite.

\smallskip
One can now use the ``smearing function" $K_I$ and the bulk-bulk kernel $G_I$ to  explicitly compute the tensor network analogue of the bulk operator reconstruction. 
In the following, we move on to interpret these results in the light of $p$-adic AdS/CFT.

\subsection{$p$-adic HKLL}
\label{sec:padicHKLL}

The  ``smearing function" $K_I$ (\ref{smearTTN}) and the bulk-bulk kernel $G_I$ (\ref{TTNbulkG}) were derived using only  ``operator pushing" of the tensor network. 
Now we show that they have clear meanings if the tree tensor network is interpreted as the bulk of  $p$-adic AdS/CFT, supporting our proposal that tree tensor network provides a concrete realization of $p$-adic AdS/CFT.

\subsubsection{Reconstruction kernel v.s.  propagator}

\smallskip
In $p$-adic AdS/CFT, the bulk propagator $g_I$ and the bulk-to-boundary propagator $k_I$ of an operator with conformal dimension $\Delta_I$ are \cite{Gubser:2016guj}:\footnote{In this subsection we will mostly focus on the case with boundary dimension  $d=1$.
}
\begin{equation}\label{padicpropagators}
g_I(v|w)=\frac{\zeta_p(2\Delta_I)}{p^{\Delta_I}}p^{-\Delta_I |\mathcal{P}(v\rightarrow w)|}
\qquad \textrm{and}\qquad k_I(v|i)=\frac{\zeta_p(2\Delta_I)}{\zeta_p(2\Delta_I-1)}p^{-\Delta_I |\mathcal{P}(v\rightarrow i)|}
\end{equation}
where the conformal dimension $\Delta_I$ is related to the mass of the scalar living on the bulk Bruhat-Tits tree by 
\begin{equation}\label{m2padic}
m_I^2=-\frac{1}{\zeta_{p}(\Delta_I-1)\zeta_{p}(-\Delta_I)}
\end{equation}
with the $p$-adic zeta function $\zeta(s)\equiv \frac{1}{1-p^{-s}}$.
\smallskip

Comparing our result of the ``smearing function" $K_I$ (\ref{smearTTN}) and the bulk-bulk kernel $G_I$ (\ref{TTNbulkG}), derived here using the ``operator pushing" for a tree tensor network,  with the boundary-to-bulk propagator $k_I$ and the bulk propagator $g_I$ of the $p$-adic AdS/CFT derived in \cite{Gubser:2016guj}, we see that\footnote{The normalizations of our smearing function and the bulk-bulk kernel will be fixed later using two-point correlation functions and will turn out to be consistent with this identification.} 
\begin{equation}\label{kernelandprop}
K_I = k_I \vert_{\Delta_I \rightarrow \sigma_I} \qquad \textrm{and}\qquad G_I = g_I \vert_{\Delta_I \rightarrow \sigma_I}.
\end{equation}

Given that the two pairs satisfy the same EOM, we should match the two expressions for the bulk mass (\ref{m2HKLL}) and (\ref{m2padic}) and obtain a relation between the parameter $\sigma_I$ in (\ref{smearTTN}) and the conformal dimension $\Delta_I$ of the operator $\mathcal{O}^I$. 
A priori, there are two solutions $\sigma_I = \Delta_I$ or  $\sigma_I = 1- \Delta_I$.

\smallskip
Recall that in the real case, the ``smearing function" (\ref{Kkernel}) used for the bulk operator reconstruction is related to the bulk-to-boundary propagator by first replacing the conformal dimension $\Delta_I$ in the exponent by $d-\Delta_I$.\footnote{The $\Theta$ function  appearing in \cite{Hamilton:2006az} is introduced as a regularization via analytic continuation of the bulk integral.} 
Therefore, the natural conclusion is
\begin{equation}\label{Deltasigma}
\sigma_I = 1- \Delta_I.
\end{equation}
\smallskip
The relation (\ref{Deltasigma})  is valid for boundary dimension $d=1$. In generic dimensions, it should be replaced by 
\begin{equation}\label{Deltasigmad}
\sigma_I = d- \Delta_I.
\end{equation}
We will prove this relation later in Section 6.

\subsubsection{Reconstruction kernel in terms of $p$-adic variables}

The relation (\ref{kernelandprop}) (together with (\ref{Deltasigma})) between the reconstruction kernels for tree tensor networks and the propagators of $p$-adic AdS/CFT shows that  the reconstruction kernels (\ref{smearTTN}) and  (\ref{TTNbulkG}) we derived using the ``operator pushing" are consistent with the expectation of the conjectured $p$-adic AdS/CFT. 
We now compare these reconstruction kernels to the HKLL formula for the real AdS/CFT and show that the tensor network provides a nice $p$-adic HKLL formula. 
\smallskip

To compare with the real HKLL, we first need to rewrite the ``smearing function" $K_I(v|i)$ (\ref{smearTTN}) in terms of $p$-adic variables. 
Interpreting the tree on which the tensor network lives as the Bruhat-Tits tree, we should assign the bulk point $v$ coordinates that label a node on the Bruhat-Tits tree as in (\ref{BTnode}). 
Namely, we start from a point on the boundary $x\in\mathbb{Q}_p$, and follow the (unique) path connecting it to the origin of the tree, and label the bulk point $v$ on the level $m$ as $(x, z=p^m)$. 
For the boundary edge $i$ in  $K_I(v|i)$, we simply assign a $p$-adic number $y$:
\begin{equation}
\textrm{bulk vertex } v: (x\in\mathbb{Q}_p,z=p^m)\qquad \textrm{and}\qquad \textrm{boundary edge } i: y\in\mathbb{Q}_p 
\end{equation}
with $m\in \mathbb{Z}$ denotes the holographic direction in the BT tree, and gives the accuracy of the $p$-adic expansion.
In particular, $|z|_p=p^{-m}$, and $|z|_p \rightarrow 0$ as one approaches the boundary. 
\smallskip

It was shown in \cite{Gubser:2016guj} that the distance from a bulk vertex on the Bruhat-Tits tree to a boundary point, in the coordinates above, is
\begin{equation}\label{distancepadic}
\vert \mathcal{P}(v\rightarrow i)\vert = -\textrm{log}_{p}\frac{|z|_{p}}{\textrm{sup}\{|z|_{p},|x-y|_{p}\}^{2}} 
\end{equation}
where the  supremum norm is $\textrm{sup}\{|z|_{p},|x-y|_{q}\}=|z|_p$ if $|z|_p  \geq |x-y|_p$ and $ |x-y|_p$ otherwise.
Using (\ref{distancepadic}) we have\footnote{We emphasize that the HKLL relation here is not to be confused with the bulk reconstruction discussed in \cite{Heydeman:2016ldy}, which is the Euclidean version of recovering the bulk field for specified Dirichlet boundary condition, which use the non-normalizable bulk-to-boundary propagator as in (\ref{padicpropagators}).}
\begin{equation}\label{padicBB}
\begin{aligned}
K_I(x,z \, \vert \, y)=
\frac{\zeta_{p}(2\Delta)}{\zeta_{p}(2\Delta-1)}
\left(\frac{|z|_{p}}{\textrm{sup}\{|z|_{p},|x-y|_{p}\}^{2}} \right)^{d-\Delta_I}
\end{aligned}
\end{equation}

Finally, recall that in the real HKLL formula one needs to regularize the bulk integration to have finite results. 
In Poincar\'e coordinates, this is done by dressing the ``smearing function" with a $\Theta((\vec{x}-\vec{y})^2-z^2)$ factor. 
In an actual tensor network computation, the tree is usually taken to be finite. 
However, the tensor network modeling $p$-adic AdS/CFT needs to live on the infinite Bruhat-Tits tree.
Therefore we propose to regularize by dressing the $p$-adic smearing function (\ref{padicBB}) with the $p$-adic analogue of the $\Theta$ function --- a factor of $ \gamma(\frac{x-y}{z})$ where $\gamma$ is the characteristic function of $\mathbb{Z}_p$ in $\mathbb{Q}_p$ defined in equation (\ref{characteristic}):
\begin{equation}\label{padicsmear}
\begin{aligned}
K_I(x,z \, \vert \, y)&=\frac{\zeta_{p}(2\Delta)}{\zeta_{p}(2\Delta-1)}
\left(\frac{|z|_{p}}{\textrm{sup}\{|z|_{p},|x-y|_{p}\}^{2}} \right)^{1-\Delta_I}\, \gamma(\frac{x-y}{z}) . 
\end{aligned}
\end{equation}

We see the $p$-adic HKLL formula, derived using the tensor network, uses a $p$-adic smearing function (\ref{padicsmear}) that is completely parallel to  the smearing function (\ref{Kkernel}) for the real HKLL formula. 
Next we show that the linear term of the bulk operator reconstruction
\begin{equation}\label{padicHKLLlinear}
\phi_p(x, z )=\int_{\mathbb{Q}_p} dy  \, K_p(\,x,z \,\vert y\,)\, \mathcal{O} (y)
\end{equation}
can be interpreted as a $p$-adic wavelet transform.

\subsection{Linear term of HKLL as wavelet transform}
\label{sec:wavelettrans}

In this subsection we show that the linear term of the bulk reconstruction can be interpreted as a wavelet transform --- a technique in signal processing, analogous to the Fourier transform but with Fourier modes replaced by the wavelet basis. 
\smallskip

As will be reviewed later, the wavelet transform has a built-in coarse graining process, therefore can be regarded as a realization of RG flow. 
For a comprehensive review of the subject, see \cite{Book}.  
This underlies the connection between wavelet transforms and AdS/CFT. 
In the context of tensor networks, it is recently found to be implementable in MERA \cite{Evenbly_White}.  
In \cite{Lee:2015vla} (see also \cite{Singh:2016mxd}) the Haar wavelet was used to construct a holographic mapping in a particular example of tensor network. 
\smallskip

We will show that the  (linear term) of the bulk reconstruction is exactly a wavelet transform, with the choice of wavelet determined universally, by AdS/CFT.  
We further show that the inverse of the bulk reconstruction is actually not the inverse wavelet transform, but needs to be regularized. 
We propose a regularization natural to the HKLL formula. 
Both the real and the $p$-adic case work in this manner.

\subsubsection{Wavelet review}
\label{waveletR}

We now review basics of wavelet transforms, using the notations in \cite{wavelet}, whose discussion makes the parallels with the HKLL relation particularly transparent. 
\smallskip

The wavelet basis containing $d+1$ parameters is defined as follows. 
First, choose the \textit{mother wavelet} $\psi (\vec{x})$, which is a (local) function of $d$ parameters $\vec{x}$.  
The set of \textit{daughter wavelets} is generated from the mother wavelet by translation by $\vec{a}$ and rescaling by $s$:
\begin{equation} \label{wavebasis0}
\psi_{\vec{a}, s} (\vec{x})\equiv \frac{1}{s^{d/2}} \psi(\frac{\vec{x}-\vec{a}}{s}),
\end{equation}
These daughter wavelets form the wavelet basis. 

Given a signal function $f(\vec{x})$, the wavelet transform using the wavelet $\psi$ is then defined as\footnote{Using the analogy with the Fourier transform,  if $\vec{x}$ is regarded as the spacetime variable, the parameter $\vec{a}$ and $s$ \textit{together} play the role of momentum variables.}
\begin{equation} \label{wavetrans}
W_f(\vec{a},s) = \int d^dx \, f(\vec{x})\, \psi^{\dagger}_{\vec{a}, s} (\vec{x})
\,, 
\end{equation}
For a generic mother wavelet $\psi(\vec{x})$, the wavelet basis $\{\psi_{\vec{a}, s} (\vec{x})\}$ it gives rise to is over-complete. 
However, special types of mother wavelet $\psi(\vec{x})$ can allow an inverse transform.
Relevant to the case at hand is when
\begin{equation} 
\hat{\psi}(\vec{k})=\hat{\psi}(k) \,,
\end{equation} 
where $\hat\psi(\vec{k})$ is the Fourier transform of $\psi(\vec{x})$ and $k \equiv |k|$. One can show that then the inverse transform of (\ref{wavetrans}) exists and is given by 
\begin{equation}\label{inverseCWT}
f(\vec{x}) = \frac{1}{C_\psi} \int_0^\infty \frac{ds}{s^{d+1}} \int d^da\, W_f(\vec{a},s)\,\psi_{\vec{a}, s} (\vec{x})
\end{equation}
where 
\begin{equation} \label{normal1}
C_{\psi} \equiv \int^{\infty}_0dk\,  \frac{|\hat\psi(k)|^2}{k} 
\end{equation}
Note that for the inverse transform to be well-defined, we also need the following ``admissibility condition"
\begin{equation}\label{admissibility}
C_{\psi} < +\infty
\end{equation}

It is often more convenient to apply the wavelet transform in Fourier space. 
Consider Fourier transforming (\ref{wavetrans}) w.r.t. $\vec{a}$. 
This gives
\begin{equation}
\hat{W}_f(\vec{k},s) =g(\vec{k},s) \hat{f}(\vec{k}) \qquad \textrm{with}\quad g(\vec{k},s)\equiv  \frac{1}{s} \hat{(\psi^\dag)}_s(-\vec{k})
\end{equation}
It means that to define the mother wavelet of a wavelet transform, we could equivalently specify a function $g(\vec{k},s)$ which is a function of the momentum and the scale factor $s$.

\subsubsection{Real linear HKLL as real wavelet}

The linear term of the bulk reconstruction (\ref{fullHKLL}) is very close to a wavelet transform with the mother wavelet 
\begin{equation}\label{mamawavelet}
\psi_{\Delta}(\vec{x})=\left(\frac{1}{1-\vec{x}^2}\right)^{d-\Delta}\Theta(1-\vec{x}^2)
\end{equation}
where $\Theta$ is the step function with $\Theta(x)=1$ for $x\geq 0$ and $\Theta(x)=0$ for $x<0$. 
Her daughter wavelets, defined via (\ref{wavebasis0}), and the HKLL ``smearing function" are related by a scaling
\begin{equation}
K(\vec{x},z \, \vert \, \vec{y}\,) = z^{\Delta-\frac{d}{2}} \psi_{\vec{x}, z} (\vec{y})
\end{equation}

We interpret this result as follows. 
The AdS/CFT correspondence selects for us a set of wavelet transform: for each operator $\mathcal{O}_{\Delta}$ of conformal dimension $\Delta$, there is a mother wavelet $\psi_{\Delta}$ such that the wavelet transform $W_{\mathcal{O}}(\vec{x},z)$, which is related to the bulk fields $\phi(\vec{x},z)$ via
\begin{equation} \label{eq:relatewavelet}
\phi(\vec{x},z) =z^{\Delta-\frac{d}{2}} W_{\mathcal{O}}(\vec{x},z)
\end{equation} 
becomes a weakly-coupled and semi-classical degree of freedom.  The holographic direction ``$z$" in Poincar\'e coordinates is identified with the scaling parameter of the wavelet transform. 
\smallskip

Now let's look at the inverse transform. 
For the standard inverse wavelet transform (\ref{inverseCWT}) to exist, the admissibility condition (\ref{admissibility}) needs to be satisfied. 
For a scalar field, the conformal dimension is $\Delta = d/2 +\nu$ with $\nu = \sqrt{d^2/4 + m^2}$. The Fourier transform of (\ref{mamawavelet})  is
\begin{equation}
\hat{\psi}_{\Delta}(k)=\frac{J_{\nu}(k )}{k^{\nu}} 
\end{equation}
which gives
\begin{equation}\label{inaddissiblereal}
C_{\psi}=\int^{\infty}_0 dk\, \frac{J_{\nu}(k)^2}{k^{2\nu+1}} \sim \Gamma(0) \rightarrow \infty
\end{equation}
i.e.\ $C_{\psi}$ diverges and the inverse wavelet transform (\ref{inverseCWT}) is no longer valid. 
However, this does not pose a problem for us because, although we use the wavelet transform to obtain the bulk field, we actually do not need to use the (standard) inverse wavelet transform  (\ref{inverseCWT})  to obtain the boundary operator from the bulk field. We can compute instead  
\begin{equation}\label{inverseHKLL}
 \int   \,d^dx \,dz\,\sqrt{g_{AdS_{d+1}}} \,\,\phi(\vec{x},z) K(\vec{x},z\vert \vec{y})\,,
  \end{equation}
which has an extra dressing factor of $z^{2\nu}$ relative to the integration measure of the (standard) inverse wavelet transform  (\ref{inverseCWT}). 
(The scaling parameter ``$s$" there is to be identified with ``$z$"  in the Poincar\'e coordinate here.) 
This is a natural choice in AdS geometry. 
 \smallskip

The Fourier transform of the smearing function is \cite{Hamilton:2006}
\begin{equation}\label{Kfourier}
 K( \vec{x},z\,\vert \, \vec{y})=
 \mathcal{N} \int_{\textrm{time-like}} \frac{d^dk}{(2 \pi)^d}  \, \frac{z^{d/2} J_{\nu}(k z)}{k^{\nu}} e^{i \vec{k} \cdot (\vec{x}-\vec{y})} 
 \end{equation}
where $k= \sqrt{-\vec{k}^2}$  and $\mathcal{N}$ the  normalization constant. 
The integral is restricted to time-like momenta so that $J_\nu$ remains normalizable. 
Plugging (\ref{Kfourier}) into (\ref{inverseHKLL}), and using 
\begin{equation}  \label{normal}
\tilde{C}_{\nu} \equiv \int_0^\infty \frac{dz}{z} \,\,J_\nu( kz)^2  = \frac{1}{2\nu}  
\end{equation} 
for time-like momenta $k$, we are left with
\begin{equation} \label{reconstruct}
\frac{\mathcal{N}^2}{2\nu}  \int_{\textrm{time-like}} \frac{d^dk}{(2\pi)^d}k^{-2\nu}  \,\, \hat{\mathcal{O}}_\partial(x')\,\,e^{i \vec{k} \cdot \vec{y}}. 
\end{equation}
Note that the integral (\ref{normal}) is the normalization that replaces $C_{\psi}$ in the wavelet transform (\ref{inverseCWT}). 
The normalization condition $\tilde{C}_{\nu}<+\infty$ in the inverse transform (\ref{inverseHKLL}) replaces the  ``admissibility condition" (\ref{admissibility}) that appears in the wavelet transform (\ref{inverseCWT}). 
The restriction to time-like momenta is due to this normalization condition, without which the integral is not finite. 
The restriction however means that the reconstruction of the original operator $\mathcal{O}_\partial$ is restricted by causality. 
Namely, equation  (\ref{reconstruct}) can be re-written as
\begin{equation}
\frac{\mathcal{N}^2}{2\nu}\int d^dx' \mathcal{O}_\partial(\vec{x}') G(\vec{x}'-\vec{y}),
\end{equation}
where 
\begin{equation}\label{realkernel}
G(\vec{x}'-\vec{y}) = \int_{\textrm{time-like}} \frac{d^dk}{(2\pi)^d} k^{-2\nu} e^{i k \cdot (\vec{y}-\vec{x}')} .
\end{equation}
This should be contrasted with the usual wavelet transform where the admissibility condition is usually satisfied for all $k$.

\subsubsection{$p$-adic linear HKLL as $p$-adic wavelet}

Just as in the real case, the linear term of the $p$-adic  bulk operator reconstruction can be regarded as the $p$-adic wavelet transform of the boundary operator.
The wavelet transform perspective of the HKLL relation can be defined in an analogous manner in the $p$-adic version of AdS/CFT.
Let's focus on the one-dimensional case. 
\smallskip

\noindent\textbf{$p$-adic wavelet transform}. 
Given a $p$-adic mother wavelet $\psi(x)$, which for our purpose is taken to be  a complex function with $p$-adic argument $x$, her $p$-adic daughter wavelets can be defined in analogue to the continuous case (\ref{wavebasis0})
\begin{equation}
\psi_{a,s}(x) =\frac{1}{\sqrt{|s|_p}} \psi(\frac{x-a}{s}) \qquad \qquad x,a,s \in \mathbb{Q}_p
\end{equation}
This set of daughter wavelets forms the basis for the $p$-adic wavelet transform. 
Then the $p$-adic wavelet transform of  a function $f(x)$ with $x\in \mathbb{Q}_p$ using this wavelet basis is
\begin{equation}\label{padicCWT}
W_f(a,s) = \int_{\mathbb{Q}_p} dx \,  \frac{1}{\sqrt{|s|_p}} \psi^\dag(\frac{x-a}{s}) f(x) 
\end{equation}
whose inverse transform is given by 
\begin{equation}\label{padicinverseCWT}
f(x)= \frac{1}{C_\psi} \int_{\mathbb{Q}^{\times}_{p}} \frac{ds}{|s|^2_p} \int_{\mathbb{Q}_p} da\, W_f(a,s)\,\psi_{a,s}(x) 
\end{equation}
with 
\begin{equation} \label{normalp}
C_{\psi} \equiv \int_{\mathbb{Q}_p} dk\,  \frac{|\hat\psi(k)|_p^2}{k} 
\end{equation}
Note that the integration measure $\frac{ds da}{|s|^2_p}$ is precisely the left-invariant measure of the $p$-adic affine group (whose action is $x\rightarrow sx+a$). 
For the inverse transform to exist, we need the same ``admissibility condition" $C_{\psi} < +\infty$, as in the real case. 
\smallskip

Just as in the real case, we can interpret the linear term of the bulk operator reconstruction
\begin{equation}
\phi_p(x, z )=\int_{\mathbb{Q}_p} dy  \, K_p(\,x,z \,\vert y\,)\, \mathcal{O} (y)
\end{equation}
as a $p$-adic wavelet transform (\ref{padicCWT}). 
\smallskip
We first simplify the ``smearing function" (\ref{padicsmear}) into 
\begin{equation}
\begin{aligned}
K_I(x,z \, \vert \, y)
=\mathcal{N}_p\, |z|^{\Delta_I-1}_p \, \gamma(\frac{x-y}{z}) \qquad \textrm{with} \quad \mathcal{N}_p\equiv \frac{\zeta_{p}(2\Delta)}{\zeta_{p}(2\Delta-1)}
\end{aligned}
\end{equation}
Now we see a  major difference from the real case. 
For the $p$-adic case, the choice of the mother wavelet does not depend on the conformal dimension $\Delta$ of the boundary operator. 
For any operator, there is a ``universal" mother wavelet 
\begin{equation}\label{padicmamawavelet}
\psi(x)=\mathcal{N}_p 
\, \gamma(x)
\end{equation}
The conformal dimension $\Delta$ only enters through the relation between her daughter wavelets and the ``smearing function":
\begin{equation}\label{padicdaughtertoK}
K_p(\, x ,z\, \vert \, y\, ) =|z|_p^{\Delta-\frac{1}{2}} \psi^{\dagger}_{x,z}(y)
\end{equation} 
which in turn gives the mapping between the wavelet transform $W_{\mathcal{O}}(x,z)$ and the bulk field:
\begin{equation}\label{padicWtophi}
\phi_p(x,z) =|z|_p^{\Delta-\frac{1}{2}} W_{\mathcal{O}}(x,z)
\end{equation} 
\smallskip

Similar to the real case in (\ref{inaddissiblereal}), our mother wavelet (\ref{padicmamawavelet}) does not obey the standard admissibility condition since
\begin{equation}
C_{\psi} =\mathcal{N}^2_p \int_{\mathbb{Q}_p} \frac{dk}{|k|_p} \gamma(k)^2 = \mathcal{N}^2_p \int_{\mathbb{Z}_p} \frac{dk}{|k|_p} \rightarrow +\infty
\end{equation}
where we have used the fact that $\gamma$ is self-dual under Fourier transform.  
Again, this does not pose a problem for us, because the physical inverse transform we use to obtain the boundary field $\mathcal{O}(y)$ is
\begin{equation}\label{padicinverseHKLL}
\int_{\mathbb{Q}^{\times}_p} \frac{dz}{|z|_p^2} \int_{\mathbb{Q}_p}dx  \, \, \phi(x,z\vert y) K_p(x,z \vert y)
\end{equation}
which has an extra dressing factor of $|z|^{2 \Delta -1}$, due to (\ref{padicdaughtertoK}) and (\ref{padicWtophi}). 
We will now show that this can make the normalization $\tilde{C}$ finite and the inverse transform (\ref{padicinverseHKLL}) valid. 
Note that similar to the real case (\ref{mamawavelet}), the function $\gamma$ restricts the integration within the ``causal patch'' in the tree. 
(In the Bruhat-Tits tree, the causal future of a node $(x,z)$ is defined as the subbranch of the tree rooted at this node.\footnote{The causal structure on the Bruhat-Tits tree was first studied in \cite{Harlow:2011az}, in the context of the hierarchical structure of inflation.})
\smallskip

First we compute the Fourier transform of the $p$-adic ``smearing" function (\ref{padicsmear}), using the additive character (\ref{addcharacter}):
\begin{equation}\label{padicsmearingFT}
K_p(\,x,z\,\vert\,y\,)=\mathcal{N}_p \, |z|^{\Delta}_p \,\int_{\mathbb{Q}_p} dk \, e^{-2\pi i [k \, (x-y)]}\, \gamma(k z)
\end{equation}
Note that since the function $\gamma$ is self-dual under the Fourier transform, in the momentum domain the restriction to time-like momenta is also implemented by the $\gamma$ function.
Plugging (\ref{padicsmearingFT}) into the inverse transform (\ref{padicinverseHKLL}) we get 
\begin{equation}
(\mathcal{N}^2_p \tilde{C}_p)\int_{\mathbb{Q}_p} dx' G_p(x'-y) 
 \mathcal{O}(x')
\end{equation}
with
\begin{equation}
G_p(x'-y) \equiv \int_{\mathbb{Q}_p} dk \, |k|_p^{-2 \nu}e^{-2 \pi i [k(x'-y)]} 
\end{equation}
the $p$-adic analogue of the kernel in (\ref{realkernel}), and\footnote{Note that the integral in $\tilde{C}_p$ only converges when $\nu>0$.}
\begin{equation}
\tilde{C}_p=\int_{\mathbb{Q}_p} dz |z|_p^{2\nu-1} \gamma(z)^2=\int_{\mathbb{Z}_p} dz |z|_p^{2\nu-1}=\frac{p-1}{p(1-p^{-2\nu})}  \qquad \textrm{for} \, \nu>0
\end{equation}
with $\nu=\Delta-\frac{1}{2}$.  Therefore, for all $\nu>0$, i.e.\ for massive scalars, the $p$-adic normalization $\tilde{C}_p$ is finite and the inverse transform is valid. 

\medskip
We have thus shown that the $p$-adic bulk reconstruction works in the same manner as the real case: the bulk reconstruction can be realized as a continuous wavelet transform, whereas the inverse, i.e. obtaining the boundary field from the bulk one, is related to the standard inverse wavelet transform by an extra dressing factor of  $z^{2\nu}$.

\smallskip
There is one important difference between the real and the $p$-adic cases. 
In the derivation of the bulk reconstruction via a $p$-adic wavelet transform, we have used the continuous wavelet transform. 
However, in essence, the $p$-adic wavelet transform is discrete.\footnote{The coincidence of the continuous and the discrete $p$-adic wavelet transform was already shown for a particular type of wavelets in \cite{Albeverio2007}.} 
The reason is the following. 
First of all, even though we started with a continuous wavelet transform in (\ref{padicCWT}), the resulting scaling parameter is $|z|_p$ instead of $z$. 
And since the integration measure in the inverse transform (\ref{padicinverseCWT}) is invariant under the affine group, the integration is actually equivalent to the sum over the Bruhat-Tits tree \cite{Gubser:2016guj}: 
\begin{equation}
\int_{\mathbb{Q}^{\times}_p} \frac{dz}{|z|_p^2} \int_{\mathbb{Q}_p}dx \, f(x,z) =\frac{1}{\zeta_p(1)} \sum_{v \in T_p}  \, f(v)
\end{equation}
Recall that the Bruhat-Tits tree can be viewed as the discrete, holographic bulk whose boundary is the continuous line of $\mathbb{Q}_p$. 
The emergence of the discrete $p$-adic wavelet transform starting from a continuous one can be viewed as the mirror statement of the above. 
This strongly supports the connection between the bulk reconstruction and the wavelet transform, with the coordinate of the holographic direction in the bulk being identified with the scaling parameter in the wavelet transform.

\section{Correlation functions and emergent Witten diagrams}
\label{correlationfn}

In this section we compute correlation functions of conformal primaries on a $p$-adic tree tensor network, and show that dual Witten diagrams emerge automatically in the bulk of the tensor network. 
We also explain the connection between the bulk operator reconstruction in Section 5 and the correlation function computation.

\subsection{Constructing conformal primaries in $p$-adic tensor network}

An $n$-point correlation function in a tensor network is defined as
\begin{equation}\label{npointdef}
\langle \psi_{\textrm{bndy}}|\mathcal{O}^{I_1} (i_1) \mathcal{O}^{I_2} (i_2) \dots \mathcal{O}^{I_n} (i_n)|\psi_{\textrm{bndy}}\rangle
\end{equation}
where the $n$ boundary operators $\{\mathcal{O}^{I_i}\}$ are sandwiched between the original ``bra" wavefunction $|\psi_{\textrm{bndy}}\rangle$ defined in (\ref{eq:holographicmap}) using the tensors $T$ and its conjugate  ``ket" $\langle \psi_{\textrm{bndy}}|$.\footnote{All other boundary edges are contracted directly between the network and its dual.}
\smallskip

To evaluate the $n$-point function (\ref{npointdef}) explicitly, we first need to construct conformal primaries $\{\mathcal{O}^{I_i}\}$. 
The conformal primary for a $p$-adic CFT was defined in (\ref{primary}). 
We have already shown that as an operator acting on the tensor network, it obeys conditions in the form of (\ref{hop}).
Now we use (\ref{hop}) to construct $\{\mathcal{O}^{I_i}\}$ using the Pauli basis (defined in (\ref{eq:genPaul})).
\smallskip
 
Just as the ``operator pushing" basis defined by diagonalizing the linear operator-pushing coefficient $\alpha$ in (\ref{HKLLemergeS}) depends on the value of the tensor $T$ (see eq. (\ref{alpha2})), the conformal primary basis also depends on $T$, though due to a different argument. 

\smallskip
Let's again use the example of the $p=2$ tree shown in Figure 
\ref{fig:treecutoff}. 
Recall that under the boundary scaling $x\rightarrow p\,x$, an operator $\mathcal{O}^I$ acting on leg 2 is mapped to $\mathcal{O}^I$ acting on leg 4. Applying (\ref{primary}) to this particular network, we have
\begin{equation}\label{hop1}
\mathcal{O}^I(i_{2})=p^{-\Delta_I} \mathcal{O}^I(i_{4}) .
\end{equation}
Above is an identify for operators on different positions. To have a local condition, we need another relation between $\mathcal{O}^I(i_{2})$ and $\mathcal{O}^I(i_{4})$.
\smallskip

The hint comes from lessons from MERA networks (shown in Figure \ref{tensornetworks}(b)), in which moving through layers of tensors realizes coarse graining. 
Since the operator is sandwiched between $|\psi_{\textrm{bndy}}\rangle$  and its conjugate  ``ket" $\langle \psi_{\textrm{bndy}}|$, a coarse graining (i.e.\ moving deeper into the tree) affects both the bulk network for $|\Psi_{\textrm{bulk}}\rangle$  and its conjugate $\langle \Psi_{\textrm{bulk}}|$.
In the $p=2$ example shown in Figure \ref{fig:treecutoff},  along the cutoff surface,
$\mathcal{O}^I(i_4)$ is one layer of tensors further away from the main branch, therefore it is related to $\mathcal{O}^I(i_2)$ as if it is sandwiched between an \textit{extra layer} of tensors. 
(This is illustrated for the case of $p=2$ in the box inside Figure \ref{fig:treecutoff}.) 
Therefore we have 
\begin{equation}\label{coarsegrain}
\mathcal{O}^I(i_{2})=T^{\phantom{*}}_{a_1 b_2 b_3} \, \mathcal{O}^I_{b_2 \tilde b_2} (i_{4}) \, T^{*}_{\tilde a_1\tilde b_2 b_3} .
\end{equation}

\begin{figure}[h!]
	\centering
	\includegraphics[trim=0.5cm 21cm 1cm 3cm, width=0.6\textwidth]{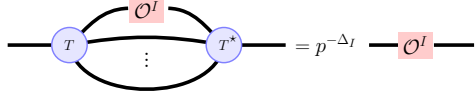}
	\caption{Illustration of equation (\ref{operator_rescaler}).    }
	\label{tensorcontract1}
\end{figure}

The two equations (\ref{hop1}) and (\ref{coarsegrain}) together give a local condition on the conformal primary $\mathcal{O}^I$:\footnote{We note that this is similar to MERA tensor networks (shown in Figure \ref{tensornetworks}(b)), in which moving through layers of tensors realizes coarse graining.
At a fixed point, the operator acting on the legs of a given layer is mapped to another ``coarse-grained'' operator by the tensor.
However, for a primary operator $\mathcal{O}^{I}$, it is simply rescaled at the fixed point, satisfying an equation like (\ref{operator_rescaler}).}
\begin{equation} \label{operator_rescale}
 T^{\phantom{*}}_{a_1 b_2 b_{3}} \, \mathcal{O}^I_{b_2 \tilde b_2} \, T^*_{\tilde a_1\tilde b_2 b_{3}} = p^{-\Delta_I} \, \mathcal{O}^{I}_{a_1 \tilde a_1}.
\end{equation}
For a generic $r=p+1$ valent tree, it is 
\begin{equation} \label{operator_rescaler}
T^{\phantom{*}}_{a_1 b_2 \cdots b_r} \, \mathcal{O}^I_{b_2 \tilde b_2} \, T^{*}_{\tilde a_1\tilde b_2 b_3\cdots b_r} = p^{-\Delta_I} \, \mathcal{O}^I_{a_1 \tilde a_1}.
\end{equation}
This is illustrated in Figure \ref{tensorcontract1}.

\begin{figure}
	\centering
	\includegraphics[trim=2.5cm 20cm 2cm 3cm, width=0.6\textwidth]{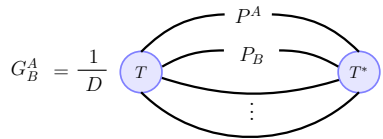}
	\caption{
	Scaling matrix $G_B^A$ for generic $r=p+1$.}
	\label{fig:gab}
\end{figure}

Finally, the conformal primaries defined in (\ref{operator_rescaler}) can be constructed starting from the Pauli basis defined in (\ref{eq:genPaul}), which satisfy
\begin{equation} \label{eq:operator_rescale2}
 T^{\phantom{*}}_{a_1 b_1 \cdots b_p} \, P^A_{b_1 \tilde b_1} \, T^*_{\tilde a_1\tilde b_1 b_2\cdots b_p} = G^A_{B} \, P^B_{a_1 \tilde a_1}.
\end{equation}
Multiplying by $P^B$ on both sides and using (\ref{eq:genPaul}), we get
\begin{equation} \label{eq:operator_rescale3}
G^A_{B} =\frac{1}{D}\, T^{\phantom{*}}_{a_1 b_1 \cdots b_p}\, P^A_{b_1 \tilde b_1} \,P^{\phantom{A}}_{B\,\,\tilde{a}_1 a_1 }
 \, T^*_{\tilde a_1\tilde b_1 b_2\cdots b_p}
  \end{equation}
as shown in Figure \ref{fig:gab}. 
Diagonalizing the matrix $G$ then gives the set of eigenvectors as the conformal primary basis $\{\mathcal{O}^I\}$, each with eigenvalue $p^{-\Delta_I}$ where $\Delta_I$ is the conformal weight. 
This gives the spectrum of the theory defined by the tensor network.

\subsection{Two-point functions}

Having constructed conformal primaries from the Pauli basis, we can now evaluate their correlation functions. 
We start with the two-point function
\begin{equation}
\langle \mathcal{O}^{I}(i)\, \mathcal{O}^{J}(j)\rangle \equiv \langle \psi_{\textrm{bndy}}|\, \mathcal{O}^{I} (i) \, \mathcal{O}^{J} (j)\,|\psi_{\textrm{bndy}}\rangle
\end{equation}
\smallskip

\begin{figure}[h!]
\centering
\includegraphics[trim=0.5cm 9cm 0.5cm 4cm, width=0.6\textwidth]{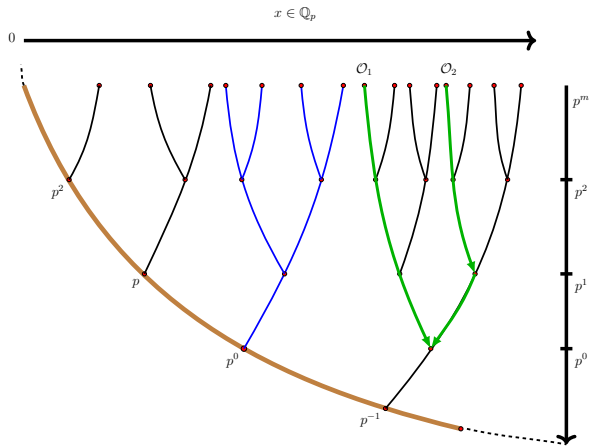}
\caption{
Pushing down the boundary operators to compute the two-point correlation function, shown for the 2-adic tree. 
}
\label{operatorontree}
\end{figure}

The computation takes two steps.\footnote{This computation is similar to computing correlators from MERA.} 
First, use the coarse graining equation (\ref{operator_rescaler}) to move both operators down, until they reach the edges of a common bulk vertex $v$, which produces a factor 
\begin{equation}
p^{-\Delta_I d(i\rightarrow v) - \Delta_J d(j\rightarrow v)}
\end{equation}
where the distance $d(i \rightarrow v)$ counts the number of edges between the boundary leg $i$ and the bulk vertex $v$. 
This step is shown in Figure \ref{operatorontree}. 

\begin{figure}[h!]
        \centering
        \includegraphics[trim=0.5cm 20cm 3cm 3cm, width=0.6\textwidth]{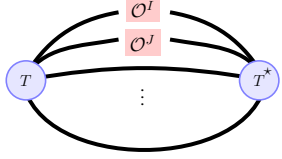}
        \caption{ Contraction in equation (\ref{eq:bulk2pt}).
        }
        \label{tensorcontract3}
\end{figure}

Then one can use the fact that the conformal primaries $\mathcal{O}^I$ and $\mathcal{O}^J$ diagonalize the scaling matrix $G$ defined in (\ref{eq:operator_rescale3}) and get 
\begin{equation} \label{eq:bulk2pt}
\langle \mathcal{O}^{I}(i)\, \mathcal{O}^{J}(j)\rangle= \delta^{I J}\, A\, p^{-\Delta_I d(i \rightarrow j)},
\end{equation}
with
\begin{equation}
 A \equiv T^{\phantom{*}}_{a_1 a_2 b_3 \cdots b_{p+1}}  \,\mathcal{O}^I_{a_1 \tilde a_1}\, \mathcal{O}^I_{a_2 \tilde a_2}\, T^*_{\tilde a_1 \tilde a_2 b_3 \cdots b_{p+1}} 
\end{equation}
as shown in Figure \ref{tensorcontract3},
and $d(i \rightarrow j)$ counts the number of edges in the (unique) path that connects the boundary leg $i$ and $j$.
\smallskip

The two point function (\ref{eq:bulk2pt}) is the result of a tensor network computation and given in terms of bulk quantities. 
Now we need to translate it into boundary $p$-adic variables. 
In the Bruhat-Tits tree, the distance  $d(i\rightarrow j)$ diverges. 
Using the regularization proposed in \cite{Heydeman:2016ldy} we have
\begin{equation}
p^{-\Delta_I d(i \rightarrow j)}\rightarrow \frac{1}{|x_i-x_j|^{2\Delta_I}_p}
\end{equation}
Remarkably, this $p+1$ valent tensor network recovers precisely the form of the two point function dictated by the SL$(2,\mathbb{Q}_p)$ invariance:\footnote{We can further normalize $\mathcal{O}^I$ such that $C_I=1$ by 
$\mathcal{O}^I (x) \rightarrow \mathcal{N}(x) O^{I} (x)$ with $\mathcal{N}(x)  \equiv  \frac{p^{\Delta_I d(x \rightarrow O)}}{\sqrt{\textrm{Tr} (\mathcal{O}^I)^2}}$ where $O$ is the origin of the Bruhat-Tits tree $(z_0, x) = (p^0,0)$.}
\begin{equation} \label{eq:2ptboundary}
\langle \mathcal{O}^{I}(x_i)\, \mathcal{O}^{J}(x_j)\rangle= \delta^{IJ}\frac{C_I}{|x_i-x_j|^{2\Delta_I}_p}.
\end{equation}
\smallskip

Finally, in this tensor network computation, as shown in  Figure \ref{operatorontree}, the two boundary operators are pushed deep into the bulk tensor network to meet at the unique bulk vertex --- a ``Witten diagram'' has emerged in the tensor network. 
The scaling relations (\ref{operator_rescale}) and (\ref{eq:cancel}) led to (\ref{eq:bulk2pt}), and the r.h.s.\ corresponds to the boundary limit of the bulk-bulk propagator which solves the scalar Klein-Gordon equation. 
The mass of the scalar is determined by the scaling dimension exactly according to the AdS/CFT dictionary.

\subsection{Higher point functions}
These considerations can be readily generalized to computing three and higher point functions, each related to an emergent Witten diagram. 
This can be compared with \cite{Vidalnpt}.
For instance, the three point vertex is given by
\begin{equation}
\lambda^I_{JK} = T_{a_1 a_2 a_3 b_4\cdots b_{p+1}} O^{I}_{ a_1 \tilde a_1} O^{\phantom{A}}_{J\,\,a_2 \tilde a_2}O^{\phantom{A}}_{ K\,\, a_3 \tilde a_3} T^*_{\tilde a_1 \tilde a_2 \tilde a_3 b_4 \cdots b_{p+1}}.
\end{equation}
They coincide with $\alpha^I_{JK}(1|2,3)$ in (\ref{eq:3ptalpha}) up to an overall normalization dependent on $D$, which can be absorbed into the definition of the reconstruction kernel. 
Having got this far, one is tempted to believe that an interacting scalar theory living in the discrete graph has emerged.

\subsection{Conformal primaries basis v.s. operator pushing basis}

A priori, it is not clear whether the primary basis obtained here coincides with the HKLL basis obtained by diagonalizing (\ref{alpha2}). 
We now show that they are actually the same basis.
Using this we can then prove the relation (\ref{Deltasigma}) in the weakly interacting limit.
\smallskip

First of all, since in the computation of the $n$-point function (\ref{npointdef}), all boundary edges --- except for the $n$ pairs that are linked by $\{\mathcal{O}^{I_i}\}$ --- are contracted directly between the  network and its dual, the identity operator should scale trivially under (\ref{operator_rescale}), which imposes a consistency condition on the tensor:  
\begin{equation} \label{eq:cancel}
 T_{a_1 b_1 \cdots b_p} \,  \, T^*_{\tilde a_1 b_1 \cdots b_p} = \delta_{a_1 \tilde a_1},
\end{equation}
\begin{figure}[h!]
        \centering
        \includegraphics[trim=0.5cm 20cm 0.5cm 3cm, width=0.5\textwidth]{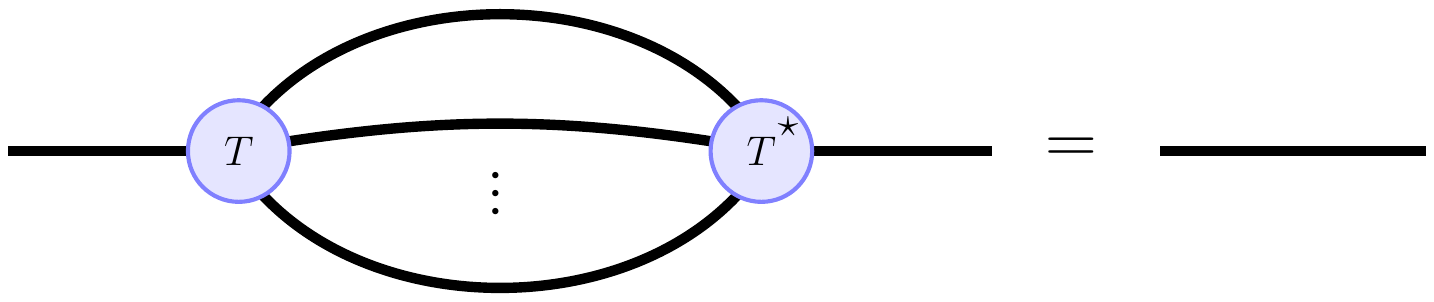}
        \caption{Consistency condition (\ref{eq:cancel}).
        }
        \label{tensorcontract2}
\end{figure}
see Figure \ref{tensorcontract2} for the example with $p=2$.
\smallskip

Recall that the tensors describing a homogenous space should be permutation invariant. 
Comparing with the $T^{-1}$ defined in (\ref{eq:weakinv}), we have
\begin{equation}\label{inversestar}
T^{-1}_{a_1\dots a_{p+1}}= T^{*}_{a_1\dots a_{p+1}}
\end{equation}
Now compare the ``operator pushing matrix" defined in (\ref{alpha2}) and shown in Figure \ref{fig:alphaab}  and the ``scaling matrix" defined in (\ref{eq:operator_rescale3}) and shown in Figure \ref{fig:gab}. 
With the identity  (\ref{inversestar}), they are proportional up to a transpose:
\begin{equation}
\alpha^{A}_B = \frac{1}{D^{p+1}}G^{B}_A
\end{equation}
Therefore the basis used in the bulk operator reconstruction and the one used in computing correlation functions are exactly the same basis. 
\smallskip

Now we prove the identity (\ref{Deltasigma}) in the weakly interacting limit (in the bulk). 
First of all, similar to the higher operator pushing coefficients as in (\ref{eq:3ptalpha}), we also need the higher scaling matrix that generalizes $G^{A}_B$ defined in (\ref{eq:operator_rescale3}). 
Sandwiching a tensor product of $p$ operators 
between $T^*$ and $T$ defines the higher scaling matrix by: 
\begin{equation} \label{eq:operator_rescale4}
 T_{a_1 b_2 \cdots b_r} \, P^{A_1}_{b_2 \tilde b_2} P^{A_2}_{b_3\tilde b_3} \cdots P^{A_p}_{b_p \tilde b_p} \, T^*_{\tilde a_1\tilde b_1 \tilde b_2\cdots \tilde b_p} =: G^{A_1\cdots A_p}_{B} \, P^B_{a_1 \tilde a_1}.
\end{equation}
Finally, inverting $P^B$ in (\ref{eq:operator_rescale4}) we get 
\begin{equation} \label{eq:operator_rescale5}
G^{A_1\cdots A_p}_{B} = \frac{1}{D} T_{a_1 b_2 \cdots b_r}\, P_{B\,\,a_1 \tilde a_1} \, P^{A_1}_{b_2 \tilde b_2} P^{A_2}_{b_3\tilde b_3} \cdots P^{A_p}_{b_p \tilde b_p} \, T^*_{\tilde a_1\tilde b_1 \tilde b_2\cdots \tilde b_p} .
\end{equation}

Similar to the operator pushing coefficients, to discuss all scaling matrices on an equal footing, we append the identity operator to the list of Pauli matrices $\{P^{I}\}$ with $I=1,\dots,D^2-1$, and define $\mathbf{1}\equiv P^{0}$. 
The scaling matrix (\ref{eq:operator_rescale5}) then includes all lower rank ones as special cases, in particular the scaling matrix (\ref{eq:operator_rescale2}) in which $p-1$ of the operators are the identity operator. 
\smallskip

\begin{figure}[h!]
        \centering
        \includegraphics[trim=2.5cm 22.5cm 4cm 3cm, width=0.8\textwidth]{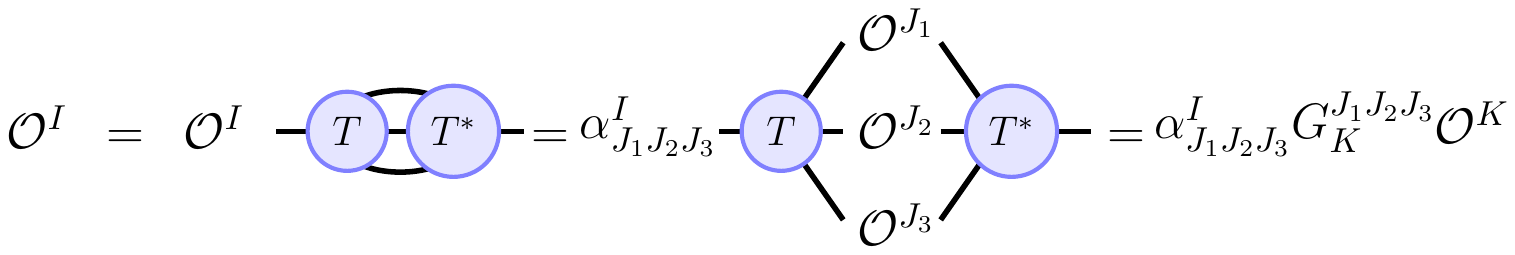}
        \caption{Chain of argument in the proof of (\ref{eq:proof}) for $p=3$.
        }
        \label{fig:proof}
\end{figure}

Consider all operator pushing coefficients and scaling matrices collectively allows us to prove a useful identity: 
\begin{equation} \label{eq:proof}
\begin{aligned}
\mathcal{O}^I_{a \tilde a} = \mathcal{O}^I_{a c} \, T_{c b_1 \cdots c_p} \,T^{*}_{\tilde a b_1 \cdots c_p} &= \alpha^I_{J_1\cdots J_p}\, T_{c b_1 \cdots b_p} \, \mathcal{O}^{J_1}_{b_1 \tilde b_1} \mathcal{O}^{J_2}_{b_2\tilde b_2} \cdots \mathcal{O}^{J_p}_{b_p \tilde b_p} T^{*}_{\tilde a_1\tilde b_1 \tilde b_2\cdots \tilde b_p} \\
&= \alpha^I_{J_1\cdots J_p} G^{J_1 \cdots J_p}_K \mathcal{O}^K_{a \tilde a},
\end{aligned}
\end{equation}
where $\mathcal{O}^{I, J, K}$ spans the list of conformal primaries (or equivalently the operator pushing basis) together with the identity matrix, i.e.\ $I, J, K=0,1,\dots D^2-1$. 
As shown in Figure \ref{fig:proof}, we have first used the identity (\ref{eq:cancel}), then the identity between $T^{-1}$ and $T^*$ (\ref{inversestar}), and finally the ``operator pushing" from one leg to the remaining $p$ legs. 

Equation (\ref{eq:proof}) then gives
\begin{equation}
\alpha^I_{J_1\cdots J_p} G^{J_1 \cdots J_p}_K = \delta^J_K.
\end{equation}
In the weakly interacting limit, where non-linear couplings $\alpha$ and $G$ involving more than two non-zero indices $I,J,K \cdots$ are suppressed relative to ``linear couplings'' $\alpha^I_J$ and $G^{I}_J$, we are left with\footnote{$d=1$ for a one dimensional boundary given by $\mathbb{Q}_p$. For higher boundary dimensions $d$, the boundary is given by $\mathbb{Q}_{p^d}$, hence each tensor in the tree carries $p^d+1$ legs \cite{Gubser:2016guj}.}
\begin{equation} \label{eq:HKvsprim}
p^d \, \alpha^I_{J} \, G^{J}_K = \delta^I_K
\end{equation}
where we have used permutation invariance of the indices of each tensor.  
\smallskip

Diagonalizing (\ref{eq:HKvsprim}) thus gives a relation between the eigenvalues $\lambda_I\equiv p^{-\sigma_I}$ of the operator pushing matrix $\alpha$,  and the eigenvalue $p^{-\Delta_I}$ of the scaling matrix $G$:
\begin{equation}
\sigma_I + \Delta_I = d
\end{equation}
which is precisely the relation between the conformal weights of the normalizable and non-normalizable solutions of the free KG equation of bulk scalar fields. 
We emphasize again that this relation emerges automatically in the tensor network.

\section{Summary and Discussion}
\label{sec:summary}

In this paper, we examined the tensor network/geometry correspondence in further detail. 
Our aim was to read off a bulk interacting theory by inspecting the properties of constituent tensors in the tensor network. 
This was done via two routes.
First by inspecting tensor identities satisfied by individual tensors, one can expand any bulk operator in terms of a sum over boundary operators. 
By organizing the sum systematically, each term acquires an interpretation as a Witten-like diagram in the bulk, in direct correspondence with the HKLL relation found in the AdS/CFT correspondence. 
This also allowed us to read off the field content in the bulk, and determine the masses of these fields and their interaction couplings.

Then we applied our methods in a specific context --- the $p$-adic AdS/CFT correspondence. 
We constructed a tensor network based on the Bruhat-Tits tree, and demonstrated that our method recovers the HKLL relation appropriate for $p$-adic AdS/CFT. 
With the bulk isometries exactly matching the boundary symmetries SL$(2,\mathbb{Q}_p)$, it implied extra physical requirements on the constituent tensors. 
These physical constraints allow us to compute boundary correlation functions exactly. 
Once again, bulk Witten diagrams emerge, from which one can read off the field content and interaction couplings. 
We demonstrated that in the ``free field'' limit in the bulk, the correlation functions follow from non-normalizable bulk-boundary propagators, as opposed to the normalizable ones that feature in the HKLL relation. 
Moreover their scaling behaviours are related in precisely the same way as in the AdS/CFT dictionary, a relation now derived based purely on properties of the tensor network. 
It gives strong quantitative support that the tensor network is an explicit embodiment of the AdS/CFT correspondence. 

Unlike the construction of MERA, we did not demand from the very beginning that $T$ should be a unitary or isometry, although a very similar condition (\ref{eq:cancel}) was later imposed for other physical requirements. 
A priori, there is no reason that the constituent tensors of a tensor network should be unitaries or isometries. 
This was imposed for efficient numerics. 
As far as a tensor network construction of wave-function is concerned, the only physically relevant condition is to provide as accurate an approximation as possible to the actual wave-function after auxiliary indices are contracted. 
A ``stochastic MERA'' where all tensors take only positive definite elements was recently constructed in \cite{Frank}.

It was noted in \cite{Heydeman:2016ldy} that the tree tensor networks do not recover  the Ryu-Takayanagi formula. 
In other words, the length of the geodesic through the Bruhat-Tits tree connecting two given end points at the boundary  in general does not give the entanglement entropy of the region bounded by the two points. 
Given that a tree tensor network in fact produces the correct correlation functions, it suggests that this is in fact a feature of  $p$-adic CFT. 
We also commented on the precise relationship between the wavelet transform and the AdS/CFT correspondence. 
Beyond its philosophical value, the relationship provides a guiding principle for future attempts at constructing  tensor networks that are dual to some target ($p$-adic) CFT's. 

Before we close our summary, we note that there is one potential confusion concerning the signature of the theory. 
In the usual discussion of the tensor network, such as the MERA and the MPSs, they are descriptions of wavefunctions, which are defined on a given time slice. 
In the $p$-adic version of AdS/CFT \cite{Gubser:2016guj,Heydeman:2016ldy}, the construction is a generalization of the Euclidean version of AdS/CFT. 
How then should the tensor network be understood in this case?  
\smallskip

To understand what happens, we recall works in the tensor network literature where classical partition functions of statistical models admit a tensor network description, see for example \cite{Gu2d, Vidal2d}.  
These tensor network representations of partition functions, which are equivalent to Euclidean path-integrals of the quantum model in one higher dimension, can also be coarse grained, which are linear maps of the constituent tensors. 
If every step of the coarse graining is kept so that those linear maps form layers of tensors in an extra dimension, we obtain a tensor network in one extra dimension. 
\smallskip

In other words, there is no mystery about the Euclidean version of the tensor network/geometry correspondence. 
It would have the same form as MERA except that what were previously physical dangling legs are now contracted among themselves in the Euclidean path-integral. 
The bulk operator boundary operator reconstruction is clearly independent of whether the boundary layers of legs are contracted among themselves.  
We acknowledge that the computation of the correlation functions as they are discussed in the current paper is rooted in the wave-function interpretation. 
There is some tension in directly interpreting the tensor network as a generating function of correlation functions. 
This however can be resolved by modifying our tensor network into a form that is more readily interpretable as a path-integral. 
We have made progress in this direction, and will report an alternative construction in a future publication.

The current work is only the first step in elucidating the relation between tensor networks and the AdS/CFT correspondence.
We list some of the imminent questions and generalizations in the following.     

\subsection{Implication to $p$-adic AdS/CFT}

One reason that $p$-adic AdS/CFT is much less developed than the real version is that there hasn't been a well-defined, but non-trivial,  ``$p$-adic CFT".  
With recent developments \cite{Gubser:2017vgc, Gubser:2017qed}, the situation is quickly evolving. 
One of the main difficulties that remain intriguing is that the ordinary derivative $\partial_z$ of the real case does not exist for the $p$-adic field $\mathbb{Q}_p$.
To proceed, there are two alternatives. 
One is to use the ``Vladimirov derivative":
\begin{equation}
D_p f(x)\equiv \int_{\mathbb{Q}_p}dy \frac{f(y)-f(x)}{|y-x|^2_p}
\end{equation}
which defines the derivative via integration, and hence can be regarded as the $p$-adic analogue of Cauchy's contour integral in complex analysis. 
The resulting field theory with a Lagrangian based on the ``Vladimirov derivative" is therefore highly non-local
\cite{Zhang:1988ku,Spokoiny:1988zk,Parisi:1988yc}.
\smallskip

The other choice is to simply  forgo the use of derivative, as in \cite{Melzer:1988he}. 
The ``$p$-adic CFT" defined this way would have only primary fields (defined using global SL$(2,\mathbb{Q}_p)$ as in (\ref{primary})), and no descendants. 
The unitary condition is then automatically satisfied and therefore cannot  impose any constraints on the conformal weights of allowed primaries. 
In fact, the form of all correlation functions is fixed and the crossing symmetry is automatically satisfied. Therefore it also cannot impose any constraints on the allowed primaries.\footnote{The modular invariance on higher genus Riemann surfaces might impose more constraints but this has not been studied.} 
\smallskip

With the result of the present paper, we can regard the tensor network living on the Bruhat-Tits tree as a concrete realization of a ``$p$-adic CFT", with different choices of the tensors corresponding to different matter content and interaction couplings. 
Since a priori, any conformal dimension is allowed, and we do not need  the derivative or Lagrangian to define the theory, this is more in line with the second, more algebraic, approach. 
\smallskip

This then allows us to go beyond the free massless scalar.  
The next simplest example would be the $\mathcal{W}_{N,k}$ minimal model,\footnote{Tensor networks have been used to study the Ising model and three-state Potts model in the real case in \cite{Gu2d}.} which also has the benefit that it is dual, by a weak/weak AdS/CFT correspondence, to a bulk higher spin gravity \cite{Gaberdiel:2010pz}. The existence of a weakly-coupled bulk dual living on the Bruhat-Tits tree then imposes additional constraints on the boundary ``$p$-adic CFT".
This is  currently under investigation.
\smallskip

\subsection{Further explorations in tensor networks}

In this paper, we have studied very specific types of graphs. 
Namely they are unweighted graphs with a fixed valency. 
Our language should admit generalizations to  more general types of graphs. 
Let us comment briefly on two interesting extensions.
\begin{itemize}
\item{For tree graphs with multiple valency, it can be described by a substitution matrix. 
A substitution matrix  $M$ is a $N\times N$ dimensional matrix, where $N$ is the number of different species, and $M_{ij}$ is the number of descendants belonging to vertices of type $j$  branching out from a vertex of type $i$ . 
For example, if we have only two kinds of vertices, we have
\begin{equation}
M=\left( \begin{array}{cc}
        a & b \\ 
        c & d
\end{array} \right)
\end{equation}
where the matrix elements are positive integers. 
One solution to the (unweighted) graph Laplacian is
\begin{equation}
G_{O}(a)= \mathcal{N} h^{m}k^{n},
\end{equation}
where $\mathcal{N}$ is some suitable  normalization and $m+n= d(a,O)$, which is the distance between a node $a$ and the {\it root} of the graph $O$.\footnote{Roughly speaking, a rooted tree contains a special point from which the tree grows outwards.} 
To satisfy the graph Laplacian equation with the same mass, we need to put in an extra constraint so that the function satisfies the same equation independently of the type of node the Laplacian acts on. 
The constraint is given by  
\begin{equation}
 c h + d k + \frac{1}{k}- (c+d+1) = a h + b k + \frac{1}{h}- (a+b+1).
 \end{equation}
It is not yet clear what such constraints could mean as far as the bulk interacting theory is concerned. It should be studied systematically.} 
\item{Another generalization is to include weighted graphs. 
In this paper, we have assumed that the network is a homogenous network and preserves as much symmetry as the graph. 
In general when we start discussing fluctuations around a background, then it seems natural that different edges should in general have different weights. 
A first step has been taken in \cite{Gubser:2016htz}, where the weights are interpreted as metric fluctuations. 
The Laplacian also depends on these weights by 
\begin{equation}
\Box\phi(v) = \sum_{u \sim v} J_{uv} (\phi(u) - \phi(v)),
\end{equation}
where $J_{uv}$ is the weight on the edge connecting the vertices $u$ and $v$.  
It is an important question to understand how these degrees of freedom emerge from the tensor network. }

\item
Time dependent evolution of the wavefunctions and dynamics in the bulk.

While a wavefunction can be evolved using any Hamiltonian that acts on the boundary legs, it is not clear whether we can interpret that evolution as some local dynamics of bulk degrees of freedom.

\end{itemize}

Apart from these generalizations, there are other important questions. 
In particular, it is evident that to recover a bulk theory that can be interpreted as a local interacting theory, there are many constraints that need to be imposed on the tensors. 
It is necessary to study these constraints systematically, and ask for the minimal set of data needed to define the bulk theory.  Such a discussion is pertinent to understanding when a large gap in conformal dimension is emerging, which is an essential condition for the emergence of a semi-classical and local holographic dual \cite{Heemskerk:2009pn}. 

\smallskip

Second,  we need a theory of gravity in the bulk, and to that end, we need to find a systematic way to describe fields carrying spin. 
As far as gravity is concerned, one way to identify fluctuations of these tensors with gravitational excitations is based on diffeomorphism. 
The idea is that the bulk graviton is supposedly dual to the boundary stress tensor. 
One could therefore ask the following question: for a given local diffeomorphism at the boundary, such as a local translation that can be effected by inserting the (exponentiated) stress tensor $e^{i \int \, dx \, \epsilon(x) T_{0x}}$, we can look for the corresponding transformation in the bulk that would have (approximately) the same effect. 
This would recover  $\delta g_{\mu\nu}$ according to the $T_{\mu\nu}$ inserted. 
The bulk transformation in general corresponds to the exchange of edges (which is particularly apparent in the case of a tree). 
It thus fits with a metric carrying two indices. 
This is currently under investigation. 

Last but not least, as we already mentioned in section \ref{sec:openq}, recovering bulk dynamics from boundary dynamics is an extremely important question. 
We hope to return to these issues in the near future.

\section*{Acknowledgements} We are grateful to Olof Ahl\'en, Sumit Das, Muxin Han, Thomas Hartman, Matthias Gaberdiel, Zheng-cheng Gu, Feng-li Lin, Hong Liu, Ezer Melzer, Mukund Rangamani, Alfred Shapere, Wei Song, Aninda Sinha, and Yichao Tian  for helpful discussions and correspondences. We especially thank Feng-li Lin for teaching us about wavelets and their recent implementation in tensor networks. 
AB thanks the Centre for Theoretical Physics at MIT, Department of Physics and Astronomy at U. of Kentucky, Physics Departments at UC Davis, UCSD, UIUC and Cornell for hospitality.  
AB and LYH thank ITP-Chinese Academy of Sciences, and WL thanks the Physics Department of Fudan University, for hospitality during various stages of this work. 
We thank support from the Thousand Young Talents Program.

\appendix

\section*{Appendix}

\section{Lattice construction of Bruhat-Tits tree}

Since $\mathbb{H}_p$ has a discrete topology, we cannot simply give $\mathbb{H}_p$ the coordinate $z_p=x_p + i\, y_p$ in which $x_p \in \mathbb{Q}_p$ and $y_p \in \mathbb{Q}_{p+}$, and write down the PGL$(2,\mathbb{Q}_p)$ invariant metric on it. 
However, the coset expression (\ref{Hcoset}) suggests that one can construct it  in terms of equivalence classes of lattices in $\mathbb{Q}_p\otimes \mathbb{Q}_p$. 
\smallskip

A lattice in $\mathbb{Q}_p\otimes \mathbb{Q}_p$ generated by basis vectors $(\vec{f},\vec{g})$ 
\begin{equation}
\vec{f} \equiv \left(\begin{matrix}f_1\\ f_2 \end{matrix}\right) \qquad         \vec{g} \equiv \left(\begin{matrix}g_1\\ g_2 \end{matrix}\right) \qquad \textrm{with} \quad f_i, g_i \in \mathbb{Q}_p
\end{equation}
which satisfy $\vec{f}\neq c\, \vec{g} $ for $\forall c \in \mathbb{Q}_p$ is defined as\footnote{Here we see an important difference between the lattices on $\mathbb{R}\otimes \mathbb{R}$ and the p-adic lattices. 
Since  $a,b\in \mathbb{Z}_p$ (the unit ball inside $\mathbb{Q}_p$), which is continuous,  the p-adic lattices  form a continuum, whereas the $\mathbb{R}\otimes \mathbb{R}$ lattices are discrete. } 
\begin{equation}
\begin{aligned}
\langle \vec{f},\vec{g} \rangle 
&\equiv  \{  a \vec{f} + b\, \vec{g} \,|\, a, b \in \mathbb{Z}_p \}.
\end{aligned}
\end{equation}
A $ \textrm{PGL}(2, \mathbb{Q}_p )$ transformation acts on the lattice via
\begin{equation}\label{pPGLlattice}
\langle \vec{f},\vec{g}\rangle  \rightarrow 
 \langle\gamma \cdot \vec{f}, \gamma \cdot \vec{g} \rangle\ \qquad \textrm{with}\qquad \gamma=\left(\begin{matrix}a&b\\ c &d\end{matrix}\right) \in \textrm{PGL}(2,\mathbb{Q}_p).
\end{equation} 
The numerator $\textrm{PGL}(2,\mathbb{Q}_p)$ in (\ref{Hcoset}) acts transitively on the space of lattices in $\mathbb{Q}_p\otimes \mathbb{Q}_p$, and the denominator $\textrm{PGL}(2,\mathbb{Z}_p)$  is the stabilizer of the integer lattice $\mathbb{Z}_p\otimes \mathbb{Z}_p$. 
Therefore $\mathbb{H}_p$ defined as coset (\ref{Hcoset}) is identical to 
the set of equivalence classes $\{\Lambda\}$ of lattices in $\mathbb{Q}_p\otimes \mathbb{Q}_p$, where two lattices are equivalent, $\langle \vec{f},\vec{g} \rangle \sim \langle \vec{f}',\vec{g}' \rangle $, iff 
\begin{equation}
        \begin{aligned}
                ( \vec{f}',\vec{g}' ) 
                =  ( \Gamma \cdot \vec{f}, \Gamma \cdot \vec{g}  ) 
                 \qquad \textrm{with}\quad \Gamma
                  \in \textrm{PGL}(2, \mathbb{Z}_p )
        \end{aligned}
\end{equation}
Note that since we use PGL instead of SL, we have
\begin{equation}\label{projective}
 \langle \vec{f},\vec{g} \rangle  \sim\lambda \langle \vec{f},\vec{g} \rangle \qquad  \qquad \textrm{with}\qquad \lambda \in \mathbb{Q}_p^{*}
\end{equation}
We denote the equivalence class of a lattice generated by $(f,g)$ as $\langle \langle f ,g\rangle \rangle$.
\medskip

\begin{figure}[h!]
        \centering
        \includegraphics[ trim=0.5cm 9cm 0.5cm 4cm, width=0.7\textwidth]{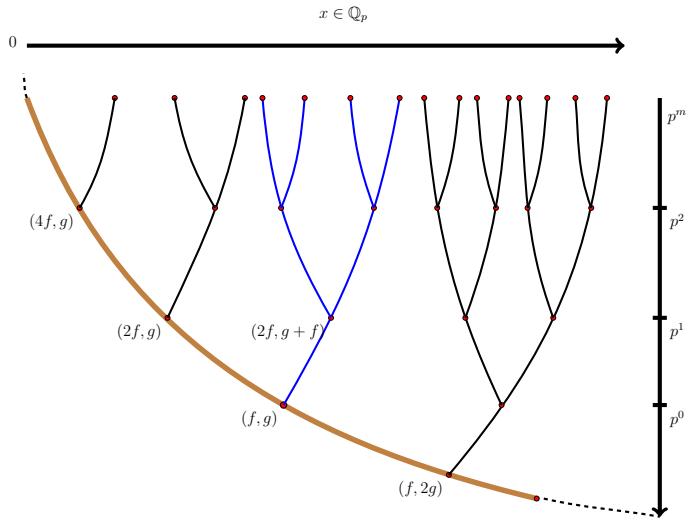}
        \caption{Bruhat-Tits tree for $p=2$. 
        Here $f \equiv \vec{f}_0$ and $g \equiv \vec{g}_0$.
        }
        \label{fig:p2bttree}
\end{figure}

Now we are ready to define the metric on $\mathbb{H}_p$, i.e.\ the distance between two points in $\mathbb{H}_p$, i.e.\ two equivalence classes of lattices $\Lambda$ and $\Lambda'$  in $\mathbb{Q}_p\otimes \mathbb{Q}_p$. 
First, $\Lambda$ and $\Lambda'$ are defined as incident (i.e.\ directly connected) if and only if\footnote{To check that this definition is reflective, we note that since $p\Lambda$ is equivalent to $\Lambda$, multiplying the equation above by $p$ gives $p \Lambda' \subset \Lambda \subset \Lambda'$.}
\begin{equation}
p \Lambda \subset \Lambda' \subset \Lambda .
\end{equation}
We connect such a pair of nodes by an edge, which has distance $d(\Lambda, \Lambda')=1$.
One can immediately check that each node has exactly $(p+1)$ nearest neighbors, i.e.\ that $\mathbb{H}_p$ has the topology of an infinite $(p+1)$-valent tree (shown in Fig.~\ref{fig:tesswithtree} for $p=2$), instead of a $\mathbb{Q}_p\otimes \mathbb{Q}_p$ continuum. 
Figure \ref{fig:p2bttree} shows the Bruhat-Tits tree for $p=2$. 
Then the distance between any two points is defined as the number of edges connecting them, which is invariant under PGL$(2,\mathbb{Q}_p)$.

For the real case, it is easy to visualize the relation between the upper half plane $\mathbb{H}$ and its boundary $\mathbb{R}$. 
In the coordinate $z=x+i\, y$ (with which the hyperbolic metric is $ds^2=\frac{1}{y^2}(dx^2+dy^2)$), taking $\lim_{y\rightarrow 0} $ gives the boundary point $x$ and increasing $y$ moves deeper in the bulk. 
An SL$(2,\mathbb{R})$ action on a point $z=x+i\, y$ on $\mathbb{H}$ would induce the same action on its boundary point $x$, see (\ref{eq:coordtrans}).
In particular, in the study of holography, the constant $y=\epsilon \ll 1$ line is taken to be the cut-off surface. 
(The choice of the cut-off surface is particularly crucial in the tensor network construction.)
What is the analogous situation for the $p$-adic case here? 
\smallskip

The lattice construction earlier only gives the topology of the tree. 
We would like to have a good coordinate system on the Bruhat-Tits tree  $\mathbb{H}_p$ such that its boundary is the $p$-adic field $\mathbb{Q}_p$, and the PGL$(2,\mathbb{Q}_p)$ action on a vertex in the bulk $\mathbb{H}_p$ induces the same action on its boundary point. 
In particular, the natural cut-off surface from this coordinate system behaves well in the tensor network construction.

\smallskip 

To fix a coordinate system in $\mathbb{H}_p$, first choose the origin $O$. 
Since the tree is infinite and homogeneous, we can choose an arbitrary point as the origin $O$. 
We can then use the isomorphism PGL$(2,\mathbb{Q}_p)$ to make $O$ the equivalence class of integer lattice, i.e.\ its generators are   
\begin{equation}
O\equiv \langle \langle \vec{f}_0, \vec{g}_0\rangle\rangle \qquad \textrm{with}\qquad         
\vec{f}_0 \equiv \left(\begin{matrix}1\\ 0 \end{matrix}\right) \qquad   
\vec{g}_0 \equiv \left(\begin{matrix}0\\ 1 \end{matrix}\right)   . 
\end{equation}
One first checks that the $(p+1)$ neighbors of $O$ are
\begin{equation}\label{nnO}
\langle \langle\left(\begin{matrix}1\\ 0 \end{matrix}\right), \left(\begin{matrix}0\\ p \end{matrix}\right)\rangle\rangle \qquad \qquad \langle \langle\left(\begin{matrix}p\\ 0 \end{matrix}\right), \left(\begin{matrix}n\\ 1 \end{matrix}\right)\rangle\rangle  \qquad \textrm{with}\quad n=0,1,\dots, p-1.
\end{equation}  
Note that these are precisely the neighbors that would arise when applying the Hecke operator 
\begin{equation}
\hat{T}_p(m) =m \cdot \left(\begin{matrix}1 &0\\ 0&p \end{matrix}\right) +m \cdot \sum^{p-1}_{n=0}\left(\begin{matrix}p &n\\ 0&1 \end{matrix}\right)
\end{equation}
on the node $\langle \langle f_0, g_0\rangle\rangle$. 
Now we use the projective equivalence (\ref{projective}) to fix the second vector $g$ to $g_0$, i.e.\ rewrite 
the first node in (\ref{nnO}) into $\langle \langle\left(\begin{matrix}\frac{1}{p}\\ 0 \end{matrix}\right), \left(\begin{matrix}0\\ 1\end{matrix}\right)\rangle\rangle
$.

Applying Hecke operator iteratively then generates the entire tree, with  all nodes having the form
\begin{equation}
        \langle \langle \left(\begin{matrix}p^m\\ 0 \end{matrix}\right), \left(\begin{matrix}x^{(m)}\\ 1\end{matrix}\right)\rangle\rangle
  \qquad \qquad x^{(m)}=\sum^{m-1}_{ n=-N} a_n p^n \qquad a_n \in\mathbb{F}_p ,  \end{equation}
where we have used the 
projective equivalence (\ref{projective}) to fix $g_2=1$. 
Figure \ref{fig:p2bttree} shows the Bruhat-Tits tree for $p=2$, with the coordinate system (\ref{Hpcoord}). 
Note that since $x^{(m)}$ truncates at $p^{m}$, we can think of $p^m$ as giving the accuracy level of a $p$-adic number $x^{(m)}$, i.e.\ the node (\ref{Hpcoord}) represents the \textit{equivalence class} $x^{(m)}+p^m \mathbb{Z}_p$.

\section{Basics of $p$-adic analysis}

In this appendix we review some basics on $p$-adic integration, $p$-adic Fourier transform, and $p$-adic wavelet transform, which will be needed when we discuss the bulk operator reconstruction in $p$-adic AdS/CFT in section \ref{sec:padicHKLL}. 
For more on $p$-adic analysis see the textbook \cite{Koblitz}. 
The summary on $p$-adic integration and Fourier transform here follows mainly the chapter 3 of   \cite{Fleig:2015vky}.

\subsection{$p$-adic integration}

The integration measure is not unique.
There are several commonly used measures for $p$-adic integration. 
First, the additive measure $dx$ over the $\mathbb{Q}_p$ line is defined by demanding the following translation and scaling behavior
\begin{equation}
d(x+a)=dx \qquad \textrm{and}\qquad d(a x)= |a|_p \, dx
\end{equation}
Usually it is normalized over the ``unit ball" (defined in (\ref{unitball}))
\begin{equation}
\int_{\mathbb{Z}_p} dx =1 
\end{equation}
Not surprisingly, the integration over the entire $\mathbb{Q}_p$ using the additive measure $dx$ diverges. 
The multiplicative measure $d^{\times}x$ over $\mathbb{Q}^{\times}_p$ can be defined by 
\begin{equation} 
d^{\times}(x) \equiv \frac{p}{p-1} \frac{dx}{|x|_p }
\end{equation}
which satisfies $d^{\times}(a x)=  \, d^{\times}x$ and $\int_{\mathbb{Z}^{\times}_p} d^{\times}x =1$. 
Finally, the Patterson-Sullivan measure $d\mu(x)$ over $\bf{P}^1(\mathbb{Q}_p)$ is defined as
\begin{equation}
d\mu(x)=\begin{cases}
dx &\qquad x \in \mathbb{Z}_p \\
\frac{dx}{|x|^2_p}&\qquad  \textrm{otherwise}
\end{cases}
\end{equation}
The distance as seen by this measure agrees with the one computed by counting steps in the BT tree.

\subsection{$p$-adic Fourier transform}
To define the $p$-adic Fourier transformation, we need the analogue of $e^{2 \pi ikx}$, i.e.\ the additive characters on $\mathbb{Q}_p$.  
This is defined as\footnote{The sign difference in the exponent between the real character and the $p$-adic one is important for constructing the adelic product.}
\begin{equation}\label{addcharacter}
 \chi_k(x)\equiv e^{-2 \pi i \,[kx]}
\end{equation}
where $[x]$ is the ``fractional part" of the $p$-adic number $x$, defined as 
\begin{equation}
[x]\equiv\sum^{-1}_{n=-N} a_{n}p^n
\end{equation}
and $[x]=0$ for $x\in\mathbb{Z}_p$, i.e. $|x|_p\leq1$ (hence the name ``fractional part"). 
It is easy to check that $\chi_k(x)\chi_k(y)=\chi_k(x+y)$ and $|\chi_{k}(x)|_p=1$. 
Integrating the additive character $\chi_k(x)$ over the unit ball  $\mathbb{Z}_p$ gives the characteristic function  $\gamma(k)$ of $\mathbb{Z}_p$ in $\mathbb{Q}_p$:
\begin{equation}\label{characteristic}
\int_{\mathbb{Z}_p} dx \, e^{-2 \pi i \,[kx]}\equiv \gamma(k) = \begin{cases} 1 \qquad k \in \mathbb{Z}_p \\ 0 \qquad \textrm{otherwise}\end{cases}
\end{equation}
Note that both the additive character $\chi_k$ and the characteristic function $\gamma(k)$ of $\mathbb{Z}_p$ in $\mathbb{Q}_p$ are locally constant functions.\footnote{As a contrast, recall that for a function from $\mathbb{R}$ to $\mathbb{R}$, ``locally constant" implies ``constant''. }

For the study of the $p$-adic Fourier transform, we summarize a few useful integrals, which are interesting in their own right. 
First, the integration over bigger balls $p^n\mathbb{Z}_p$ gives the characteristic function of $p^n\mathbb{Z}_p$ in $\mathbb{Q}_p$, i.e. $\gamma(p^nk)$:
\begin{equation}
\int_{p^n\mathbb{Z}_p} dx \, e^{-2 \pi i \,[kx]} =\frac{1}{p^n}\gamma(p^nk)
\end{equation}
From this we can compute the integration over ``shells"
\begin{equation}
\int_{p^n\mathbb{U}_p} dx \, e^{-2 \pi i \,[kx]} = 
\begin{cases}
\frac{p-1}{p^{n+1}} \qquad \qquad & |k|_p < p^{n+1}\\
\frac{-1}{p^{n+1}}\qquad \qquad & |k|_p = p^{n+1}\\
0 \qquad \qquad &|k|_p > p^{n+1}
\end{cases}
\end{equation}
where $\mathbb{U}_p$ is the ``unit sphere" of $p$-adic numbers defined in (\ref{unitsphere}). 
This  in turn gives
\begin{equation}
\int_{\mathbb{Q}_p \backslash\mathbb{Z}_p} dx \, e^{-2 \pi i \,[kx]} =- \gamma(k)
\end{equation}
\smallskip

The $p$-adic functions that allow a well-defined (i.e.\ invertible) Fourier transform are locally constant (i.e.\ constant within each $p^n\mathbb{Z}^{\times}_p$) functions with compact support or sufficiently fast asymptotic decay.
The $p$-adic Fourier transform and its inverse is then 
\begin{equation}
\hat{f}(k)=\int_{\mathbb{Q}_{p}} dx \, f(x) e^{-2\pi i[kx]} \qquad\textrm{and} \qquad  f(x)=\int_{\mathbb{Q}_{p}} dk\, \hat{f}(k) e^{2\pi i[kx]}
\end{equation}
\smallskip

The characteristic function $\gamma(x)$ of $\mathbb{Z}_p$ in $\mathbb{Q}_p$ is invariant under the Fourier transform, i.e.
\begin{equation}\label{padicgammaFT}
\hat{\gamma}(x)=\gamma(x)
\end{equation}
hence $\gamma(x)$ is also called $p$-adic Gaussian.

\section{Some examples}

\subsection{GHZ tensor}

One very simple example that satisfies permutation invariance and (\ref{eq:cancel}) at the same time is given by the GHZ state, which can be defined for arbitrary valency $r$ and bond dimension $D$. 
The only non-vanishing components are given by
\be
T_{a\cdots a} = 1, \qquad1 \leq a \leq D .
\ee

For simplicity, let us consider $D=2$ and $r=3$ explicitly. There, the Pauli basis is simply given by the Pauli matrices $\sigma_{x,y,z}$. Therefore we can check that only $\sigma_z$ has a linear term under operator pushing. The linear pushing coefficient defined in (\ref{alpha2}) and the scaling relation defined in (\ref{eq:operator_rescale2}) are given by
\be
\alpha^Z_{Z} = \frac{1}{2} = \frac{G^Z_Z}{2} .
\ee
This means that $\sigma_z$ has scaling dimension $\Delta=0$. In this case, the cubic coupling $\alpha^A_{BC}$ and $G^{AB}_C$ vanish. We can then explicitly see that $2 \alpha^Z_Z \times G^Z_Z = 1$, satisfying (\ref{eq:HKvsprim}). 

The other two Pauli operators $\sigma_{x,y}$ however, have infinite scaling dimension. 
The GHZ state for general dimensions generates unfortunately either operators with vanishing or infinite scaling dimensions. 
Despite not having a physical spectrum, it does illustrate in very simple terms the ideas described in this paper. 

\subsection{Random tensors}

Consider the coupling constants in a generic tensor network built from $r$ leg tensors with large bond dimensions $D$. 
We can use Haar weighted random tensors to estimate how the non-linear couplings scale with $D$ in the large $D$ limit.

First, we compute two point functions to appropriately normalize the operators.
From the requirement that (\ref{eq:cancel}), we require
\begin{equation}
\mathcal{N}^4 \overline{\textrm{tr}(|T\rangle\langle T | \otimes |T\rangle \langle T| )} = D.
\end{equation}
Now using results from \cite{Hayden:2016cfa}, we have
\begin{equation}
\overline{\textrm{tr}(|T\rangle\langle T | \otimes |T\rangle \langle T| )} = \textrm{tr} \frac{I + \mathcal{F}}{D^{2r} + D^r}  = 1,
\end{equation}
where $\mathcal{F}$ is the swap operator swapping the two copies  i.e. $\mathcal{F}|T_1\rangle \otimes T_2\rangle =|T_2\rangle \otimes T_1\rangle$.
Therefore $\mathcal{N}=\sqrt{D}$.

Then 
\be
\overline{(G^A_B)^2}= \frac{D^2}{D^2} \frac{\textrm{tr}(({P^{A}}^2)   \textrm{tr}(({P^{B}}^2)(\textrm{tr}(1)^{D-2}) }{D^{2r}+D^r} \sim \frac{1}{D^r}.
\ee
On the other hand, three point couplings are given by exactly the same value. i.e.\
\be
\overline{(G^A_{BC})^2} = \frac{1}{D^r}.
\ee
We therefore have 
\be
\frac{\overline{(G^{A}_{BC})^2}}{(\overline{{G^A_B}^2})^{3/2}}  = \frac{D^{-r}}{D^{-3r/2} }= D^{r/2},
\ee
which diverges in the large $D$ limit, as expected.

\section{Bulk operator reconstruction in the vicinity of the perfect code} 

In this appendix, we will demonstrate how the discussion of HKLL-like relations and fusion matrices in equation (\ref{bulk0}) are related to the discussion in \cite{Bhattacharyya:2016hbx}. 
To that end, like \cite{Bhattacharyya:2016hbx}, we will have to work in the limit where every tensor is close to being a perfect tensor. Recall that a perfect tensor is one satisfying
(\ref{eq:perfectdef}). Now we will consider tensors at each site taking the form
\be
\hat{T} = T + \epsilon \, t\, ,
\ee
where $T$ is a perfect tensor, and $t$ is some arbitrary tensor and $\epsilon$ an infinitesimal parameter. 

\begin{figure}[H]
        \centering
      \includegraphics[trim=0.5cm 20cm 5cm 3cm, width=0.5\textwidth]{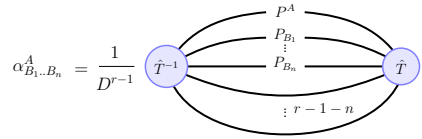}
        \caption{ This depicts how the expansion coefficients  $\alpha^A_{B_1\cdots B_n}$ can be computed.     }
        \label{fig:melon}
\end{figure}

We would like to extract the index $\alpha^A_{B_1\cdots B_n}$. 
This can be obtained by multiplying both sides by the set of out operators $\otimes_i P^{B_i}$, and then taking the trace. This is depicted in Figure \ref{fig:melon}.

To order $\epsilon^0$, $T^{-1}$ is indeed just $T^*$. Using (\ref{eq:perfectdef}) and (\ref{eq:genPaul}), one can readily show that the first coefficient that would show up at $\epsilon^0$ has to be $\alpha^A_{B_1,\cdots B_{r/2}}$. This is indeed the case in every explicit example of a perfect tensor.

\section{Boundary global Space (time) symmetry via tensor transformations }\label{sec:TNsymmetry}

In this appendix, we would like to make explicit how the spacetime symmetries are realized via tensor transformations. 
We note that the tensor network has so far only been used as a description of wavefunctions. 
Therefore symmetries related to boosts are not explicitly respected. 
However, if we focus on space-like symmetries, or take the tensor network as a Euclidean bulk, then the statement of symmetries can be taken as follows.

For a state invariant under some space-symmetry transformation $g$, it means
\begin{equation} \label{ginvariance}
g | \Psi\rangle = |\Psi\rangle.
\end{equation}
For example, $g$ could correspond to quantum operators implementing translations, rotation or scaling in a CFT.
The state $|\Psi\rangle$ however is constructed from contraction of tensors in a network. 
This means that the description of the state in terms of these tensors involves a large amount of redundancy --- any contracted leg between two neighboring tensors $T^I$ and $T^J$ can be rotated by a unitary transformation:\footnote{We use capital Latin letters to label tensors on individual sites of the network, Greek letters for contracted indices, and small Latin letter for boundary legs.}
\begin{equation}\label{gauge}
\cdots T^I_{a\, \alpha}T^J_{ \alpha\, b}\cdots = \cdots T^I_{a \, \alpha}U_{\alpha \beta} U^{\dag}_{\beta \gamma}T^J_{\gamma \, b}\cdots .
\end{equation}
This transformation $U$ is often also referred to as a gauge symmetry in the standard tensor network literature. 
\smallskip

Therefore the symmetry transformation is defined up to a gauge transformation
\begin{equation} \label{eq:Ttrans}
g_{ab} T(v)_{b,i_{e_1}',i_{e_2}',\cdots} = T(\hat{g}.v)_{a, i_{e_1},i_{e_2},\cdots} W^g_{i_{e_1},i_{e_1}'}(e_1) W^g_{i_{e_2},i_{e_2}'}(e_2) \cdots.
\end{equation}
Here we made the distinction between the symmetry transformation of the CFT wavefunction, denoted $g$, and the geometric transformation $\hat{g}$ of the tensor network graph. 
In a holographic type tensor network, this is very interesting. 
In addition to inducing gauge transformations, the tensor at the top level has moved elsewhere. 
It means that  the tensor on the RHS of the equation has to move to a different vertex $\hat{g}.v$, which is precisely one dictated by the action of the transitive action of the symmetry in the bulk. 
In the case of the Bruhat-Tits tree, $\hat{g}.v$ would be defined by equation (\ref{eq:treetrans}).

\smallskip

The above equation only considers the simple situation in which motion of the vertices in the bulk suffices. 
This is the case if we impose also permutation invariance of all the legs of an individual tensor. 
In general, the symmetry transformation might also involve the permutation between legs. 
This would require a systematic study. 
In the following, we will only briefly illustrate this  proposal in a 3-qutrit code discussed also in \cite{Pastawski:2015qua}. 

It is interesting to contrast space-symmetries with global internal symmetries. 
In the latter, it is known in the literature that it is convenient to impose invariance of individual tensors under the action of the global symmetry group. 
As a result, the global symmetry is enhanced to a gauge symmetry in the bulk, precisely as expected in the AdS/CFT dictionary. 
Here, the ``gauge transformations'' appearing in the RHS of (\ref{eq:Ttrans}) are probably related to the bulk diffeomorphism. 
But it probably does not recover the full scope of diffeo-invariance of the bulk. 
A proper description would amount to recovering all the redundancy in the tensor network, and subsequently the closest analogue of gravity in the discrete space-time. This is certainly beyond the scope of the current paper, and we will restrict our attention to global bulk isometries. 

\subsection{Example using the 3-qutrit code}
We now study the example of the pentagon code. 
We will consider two different types of wavefunctions to illustrate the above point. 
For concreteness, one can take the 3-qutrit code \cite{Pastawski:2015qua}.  
Then each tensor $T^I_{\alpha\beta\gamma}$ has 4 indices, each taking 3 values $\{0,1,2\}$.  
The index $I$  is treated as a bulk index. 
To illustrate how the symmetry acts, we can take these tensors to populate the 2-adic tree.

Consider for simplicity performing a scaling transformation. 
As described in section \ref{sec:scaling}, this corresponds to a translation of branches along the main branch, see figure \ref{tess2}.  
Therefore, a scaling transformation of the wave function would correspond to moving the tensors in exactly the same fashion as the nodes of the tree, while preserving the orientation of each tensor (if they are not already permutation invariant). 

Practically, to effect a translation in this tree, we can make use of the following operator:
\begin{equation} \label{permOP}
\mathcal{T}=\prod_{l, <v_{i_l},v_{i_l+1}>}(I^{v_{i_l}} \otimes I^{v_{i_l+1}}+ \sum_{M=1}^{D^2-1} (P_M^{v_l} \otimes P_M^{v_{i_l+1}})),
\end{equation}
where the superscript $v$ denotes the site index. $P^{v_{i_l}}_M$ are the generalized Pauli matrices as described in (\ref{eq:genPaul}) and $I^v$ is the identity matrix that acts on the bulk index of the tensor located at vertex $v$. 
For the 3- qutrit code therefore, $D=3$. 
Each term in the product exchanges the $v_{i_l}$ site and the $v_{i_l+1}$ site. 
The product runs over the sequence of vertices, i.e.\ the set of vertices $\{v_{i_l}\}$ which transforms as $v_{i_l}\to v_{i_l+1}$, and where $l$ is a ``layer'' index. 
Translation is effected within each layer $l$.   
The wave function generated by this  transformation is one in which the boundary sites are transformed according to $x\to p x$.

\bibliographystyle{JHEP}
\end{document}